# uStash: a Novel Mobile Content Delivery System for Improving User QoE in Public Transport

Fang-Zhou Jiang, Kanchana Thilakarathna, Sirine Mrabet, Mohamed Ali Kaafar, and Aruna Seneviratne

**Abstract**—Mobile data traffic is growing exponentially and it is even more challenging to distribute content efficiently while users are "on the move" such as in public transport. The use of mobile devices for accessing content (e.g. videos) while commuting are both expensive and unreliable, although it is becoming common practice worldwide. Leveraging on the spatial and temporal correlation of content popularity and users' diverse network connectivity, we propose a novel content distribution system, *uStash*, which guarantees better *QoE* with regards to access delays and cost of usage. The proposed collaborative download and content stashing schemes provide the *uStash* provider the flexibility to control the cost of content access via cellular networks. We model the *uStash* system in a probabilistic framework and thereby analytically derive the optimal portions for collaborative downloading. Then, we validate the proposed models using real-life trace driven simulations. In particular, we use dataset from 22 inter-city buses running on 6 different routes and from a mobile *VoD* service provider to show that *uStash* reduces the cost of monthly cellular data by approximately 50% and the expected delay for content access by 60% compared to content downloaded via users' cellular network connections.

**Index Terms**—Crowd-sourced, Public Transport, Spatial and Temporal, Cooperative Networking, Last mile, QoE.

## 1 Introduction

The pervasive use of smart mobile devices has led to an exponential growth of cellular data traffic. Network operators often introduce capped data plans which leads to increased costs experienced by mobile users as well as unsatisfactory quality of experience (*QoE*). The predictions are that the demand will outpace the increases in capacity that will be offered by new technologies such as LTE+ or 5G. The low *QoE* issue becomes even more acute for commuters on public transportation systems due to the poor connectivity induced by high mobility and concentration of users [1]. The approaches taken for improving *QoE* of commuters such as pre-fetching and caching [2], [3] or having parallel multiple cellular links [4], have not been so far widely adopted mainly due to the difficulties of predicting user behavior and the associated cost for service providers.

Currently, commuters could choose to utilize their own cellular connections or a local network provided by public transport provider when available, which is connected to the Internet via a cellular back-haul link. For the first approach, commuters suffer from the problems alluded to above, and the bandwidth usage contributes directly to their mobile data plans. There have been researches trying to keep users best connected by providing better service in terms of coverage and hand-over [5], while the cost for accessing content via cellular network is high and the *QoE* is limited. With regards to the second approach, many service providers around the world have been providing free Internet access on-broad public transport [6], [7], although the number of provider is low and coverage is still very limited. The problem for this approach could be noticed especially when the vehicle is moving in and out of different base stations [8]. Furthermore, a bus or train full of mobile users could easily overwhelm the capacity of the mobile access point and the cellular back-haul link. The *QoE* could be even worse as the cellular links are normally constrained to a single service provider and the cost for bus companies in providing the data services could be profound.

In this paper, we propose an alternative approach that consists of exploiting the transient co-location of mobile users and potential spatio-temporal correlation of content popularity along with the high capabilities of the modern mobile devices. We propose *uStash*, the rationale for which was inspired by the delivery of information via the free newspapers in public transportation systems, where users consume content (reading the paper) whilst traveling but leave the content (the paper) when they leave. In *uStash*, when a user requests content, the system checks its availability in its storage (*stash*) via the local network. If there is a stash hit, the content is delivered via the local network. If there is a stash miss, the content is cooperatively downloaded by the user and the stash. The portion of the content that the user needs to download is dynamically determined by the stash. Once a user finishes watching (consuming) the downloaded content, the user downloaded portion is pushed to the stash for use by others.

This is fundamentally different from an in-network cache system, as the stash is not in the data path and content is pushed to the stash via a different network from which it was downloaded. As different users have subscriptions with different network operators, this approach enables the use of diverse network connections and thus maximizes availability and robustness. Moreover, users can reduce their costs and the *uStash* providers can limit their expenditure by specifying the portion of content that will be downloaded by

- *F. Jiang, K. Thilakarathna, S. Mrabet, M.A Kaafar and A. Seneviratneare are with the Data61, CSIRO, Australia, E-mail: firstname.lastname@data61.csiro.au.*
- *F. Jiang, and A. Seneviratne are also with School of EE&T, University of New South Wales. K. Thilakarathna is also affiliated with School of Information Technologies, University of Sydney* them. With *uStash* there is no need to predict users' interests as what is available in the stash will be what other users of the system consumed. And finally there is no requirement for users to give access to their devices to strangers. The appropriateness, content integrity, and authenticity can be perceived as an obstacle for *uStash* deployment. However, these concerns can be adequately addressed by well-known techniques that are used in the user generated content hosting platforms, i.e. *YouTube*. This paper makes the following contributions;

- With the support of large-scale real-world datasets, we postulate and verify possible relationships between the degree of geographic locality and granularity, extending the current findings to mobile edge.
- We propose a novel system, *uStash* that improves *QoE* and reduces the cost of data download for commuters, as well as the network access provider of public transportation systems.
- We model the *uStash* system and derive the optimal portion to be downloaded by the stash to satisfy the objectives of minimum expected completion time, costs for both users and the stash provider. Additionally, the system model is validated with real-life data driven simulation.
- We show that stashing partially downloaded segments can be very beneficial with partial hits of more than 50% of total video content hits.
- We analytically show that by using *uStash* delay of downloads can be reduced by 25% when compared to using WiFi hotspots and 60% when compared to direct cellular network downloads.
- We demonstrate that the required stash size is well within the storage capabilities of a small device such as a *Raspberry-Pi* and show the practical viability of *uStash* by implementing it on a *Raspberry-Pi* for *Android* users.

The rest of the paper is organized as follows; Section 2 provides the motivation for *uStash*, and describes the datasets used. The system description and modeling of *uStash* are presented in Section 3. *uStash* performance is evaluated via simulation in Section 4, followed by a prototype implementation and "in the wild" experiment in Section 5. Section 6 summarizes the related work. Section 7 provides the conclusions.

## 2 Motivation

**Rationale for on-board uStash**: We postulate and verify that commuters on a public transport vehicle, such as a bus or a train carriage at a given point of time, on a given day, are more likely to be interested in similar content. For example, a group of travelers to a tourist attraction would have more similar interests than a random group of people, and would have quite different interests to a group of commuters traveling from a city's financial district after work. In addition, regular commuters are likely to access similar content periodically, i.e. the local newspaper on a daily basis during the working week. In other words, there is a spatial and temporal correlation with the content that is requested by the users when commuting. The possible existence of spatio-temporal correlation of commuter interest motivates us to further push caching functionality of network paradigm across the "last mile" to mobile user proximity as a stronger correlation in content interests translates to a higher probability of edge stash hit rate.

However, the functionality of uStash does not purely reply on stash hits, and it is designed to outperform current existing systems with regards to user *QoE* even in case of stash misses. Since commuters have subscriptions with different cellular network operators, the user group on a public transport vehicle as a whole will have diverse paths to access content. Such diversity of paths, along with the spatio-temporal correlation of the data being requested can be leveraged to provide a more efficient and robust content delivery service to users of public transportation systems. We use the datasets described in the next section to validate our assertion of spatio-temporal correlation of content access by the commuters.

### 2.1 Datasets in use

*The On-board WiFi Dataset:* The system consists of an on-board gateway device installed on bus that provides a local WiFi network. The gateway device provides Internet access by transparently redirecting all traffic to a land-based *Squid* proxy via a cellular network. The data we use was extracted from the *Squid* logs for 5 weeks from February to March, 2015. The dataset contains filtered HTTP header information of passenger mobile devices from 22 inter-city buses (referred as bus lines #5561-5582) on 6 different routes. Overall, there are 1,140,757 unique content requests for 26,699 unique domain names. We refer to this dataset as *DS1*. Table 1 summarizes the basic statistics of the dataset.

*The Online VoD Service Dataset:* This dataset consists of server logs from one of the largest *Video On Demand* services in the world. The logs were collected in December 2014, and contain randomly sampled content requests from three cities, referred to as City 1, 2 and 3, corresponding to 569,555 requests, generated by more than 420,000 mobile users. We refer to this dataset as *DS2*. The users in *DS1* account for content requests of 147GB in size and *DS2* users request 70TB of mobile video content. *DS2* consists of the viewed length of the video in addition to the total length of the video, with a watch ratio of 25% on average.

### 2.2 Validation of Assertion

We first validate the postulation by confirming the existence of spatio-temporal correlation of content access using both *DS1* and *DS2*. Let $\mathbb{D}_i$ be the set of content items requested by bus $i$. We then compute the **Jaccard similarity**, $JS(i,j) = \frac{\|\mathbb{D}_i \cap \mathbb{D}_j\|}{\|\mathbb{D}_i \cup \mathbb{D}_j\|}$ for two given buses $i$ and $j$. Fig. 1a depicts the inter-bus similarity with **JS** values among all pairs of buses. Interestingly, the similarity among the buses of the same route are significantly higher than inter-route buses. For instance, *all* pairs of buses on Route 1 show **JS** values higher than 27.5%. This confirms the existence of a spatial correlation of content access. Therefore, mobile commuters traveling on the same route, have significantly similar interests compared to inter-route commuters, thus confirms the existence of spatial correlation of content access.

TABLE 1
Summary of the on-board WiFi – DS1

| Route | City A (Pop.) | City B (Pop.) | # Requests/Bus 1 | 2 | 3 | 4 | Distance |
|---|---|---|---|---|---|---|---|
| 1 | 789k | 65k | 222,827 | 294,998 | 181,039 | 193,265 | 36km |
| 2 | 30k | 65k | 119,838 | 138,432 | 53,962 | 193,002 | 84km |
| 3 | 130k | 87k | 90,044 | 45,556 | 68,549 | 96,508 | 111km |
| 4 | 87k | 27k | 304,680 | 211,437 | 241,787 | 230,429 | 67km |
| 5 | 87k | 18k | 111,097 | 103,294 | 129,023 | | 43km |
| 6 | 87k | 1k | 115,383 | 32,976 | 89,126 | | 120km |

Similarly, we calculate the **JS** values for content requests between days. Fig. 1b illustrates the existence of temporal correlation of content access. In specific, we show that the daily similarity of popular content decreases in time, while there is a periodic pattern of high similarity between content categories accessed on the same days of the week. Weekends also exhibit a very different work-load/content set compared to weekdays. Moreover, Friday has the highest **JS** value with the following Monday rather than the weekend. This is most likely due to the cohort of commuters on weekends being different (typically non-returning) to the commuters during the weekdays (typically returning commuters). In addition, the closer the days, the higher the similarity, which is exemplified by the relatively higher **JS** values of the lower left triangle than the top right, where the difference between the two days are less than 7 days. These user behaviors validate the existence of a temporal correlation of content access.

We then calculate the entropy $E_S(i)$ of a content request $i$ among different routes and among the buses in the same route, using Equation 1. To remove the bias towards highly popular route/bus, the popularity of the request $\alpha_i^s$ at the particular route/bus is considered instead of the absolute number of requests.

$$E_S(i) = -\sum_{s=1}^{n} \frac{(P_i^t \times \log P_i^t)}{\log n}, \text{where } P_i^t = \frac{\alpha_i^s}{\sum_{s=1}^{n} \alpha_i^s} \quad (1)$$

$P_i^t$ is the probability of request $i$ coming from the route/bus $s$. The CDF of $E_S(i)$ in Figure 2 indicates that 40% of the content requests originated from a single route and approximately 60% of those requests generated from a single bus within the route. Moreover, mean intra-route entropy is 10% higher than mean inter-route entropy, suggesting a concentration of interest in a finer geographical granularity. This surprising result further justify the strong spatial locality of user interests.

Despite the high spatial locality of user interest, preliminary results of caching content at the bus with DS1 suggests that a cache hit rate of over 20% on average can be achieved [9], which is more than twice as high as caching at the cellular base stations as shown in [10]. The difference in hit rate, we believe, lies in the fact that a particular bus is small enough to carry a set of users with similar interest, contradicting the common believe that caching performance is poor at mobile edge. This observation strongly supports our proposal of stashing content at buses and trains.

Furthermore, our postulation of strong correlation at mobile edge is also supported by the general trend that geographic locality is strong everywhere [11], [12] and becoming stronger at a finer geographic granularity. In Table 2, we listed relevant studies in the field of geographic locality (spatial correlation) backed by large-scale datasets, and plotted the general trend in Fig. 3 with two possible relationships. Relationship 1 is the lest squares exponential fitting all points at different granularity, and Relationship 2 assumes no change for granularity coarser than city-level and shows the lest squares exponential fitting for points finer than city-level. Moreover, although geographic locality is quite strong for granularity coarser than city level, it is particular strong for finer granularity areas regardless of the actual relationship. More data of similar nature needs to be collected for validating the change of geographic locality with respect to granularity, however, a clear increase at mobile edge is observed. Our finding extends the state-of-the-art geographic locality studies to finer geographical granularity (mobile edge).

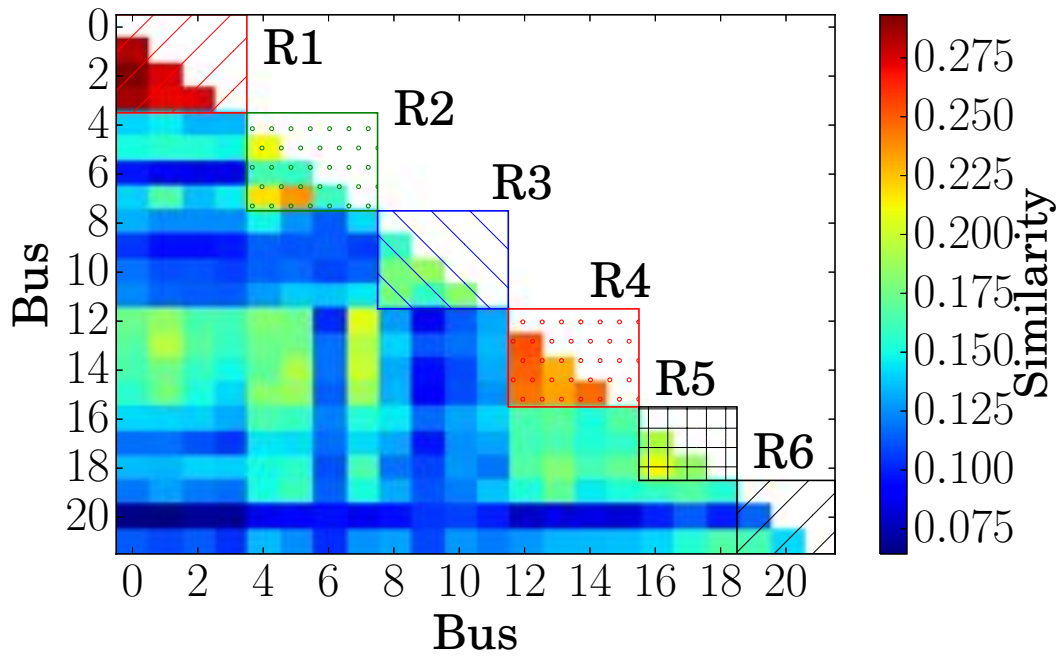

(a) Spatial similarity

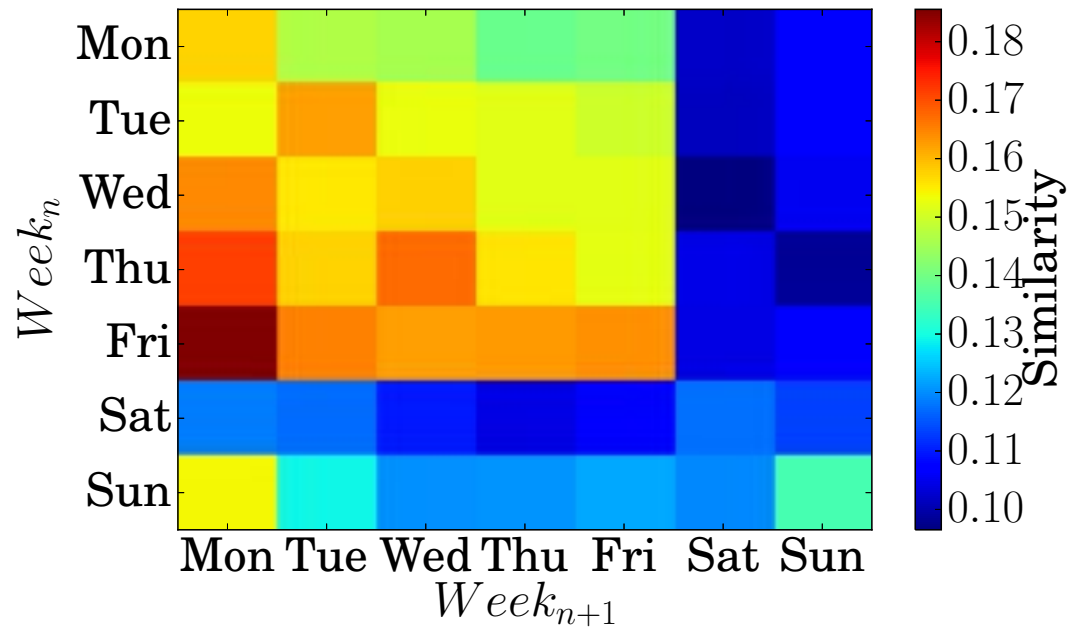

(b) Temporal similarity

Fig. 1. Spatio-temporal correlation

Next, we introduce the *uStash* system for public transport, which leverages the observed spatio-temporal correlation, and the inherent path diversity to improve *QoE* and reduce costs of providing access to popular content for commuters. The proposed system is easy and cheap to install, while saves cellular data cost for public transport companies in the long run. *uStash* is designed to save bandwidth cost for both service provider and users and at the same time improve users' *QoE*. Our postulation of spatio-temporal correlation offers more incentive for the case of mobile edge caching, while the design of *uStash* enables superior performance compared to current systems even when there is a stash miss.

TABLE 2
Validation of Spatio-Temporal Correlation

| Geo Granularity | Dataset | Geographic locality | Uniqueness |
|---|---|---|---|
| Inter-Regional<br>Inter-Country | 20M YouTube [11]<br>650K YouTube [13] | Quite Strong | 18% of videos have all views from a single region |
| Inter-Provincial | 86M VoD log [12] | Quite Strong | 20% of videos have all views from a single province |
| Inter-City | 9.6M VoD log [14], *DS2* | Quite Strong | 18% of videos have all views from a single city |
| Inter-Route Bus | *DS1* | Very Strong | 40% of content are originated from a single route |
| Intra-Route Bus | *DS1* | Strongest | 60% of content generated from a single bus |

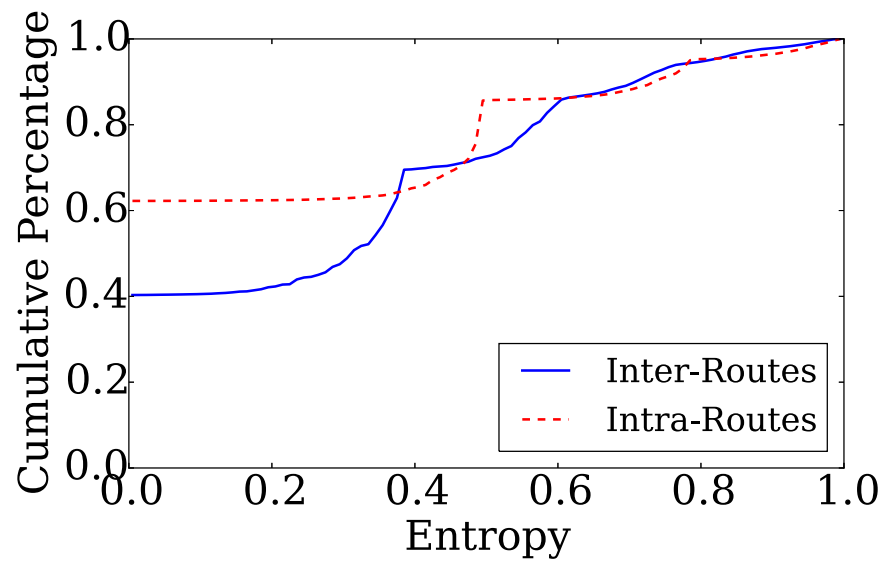

Fig. 2. Intra-bus and inter-bus Entropy

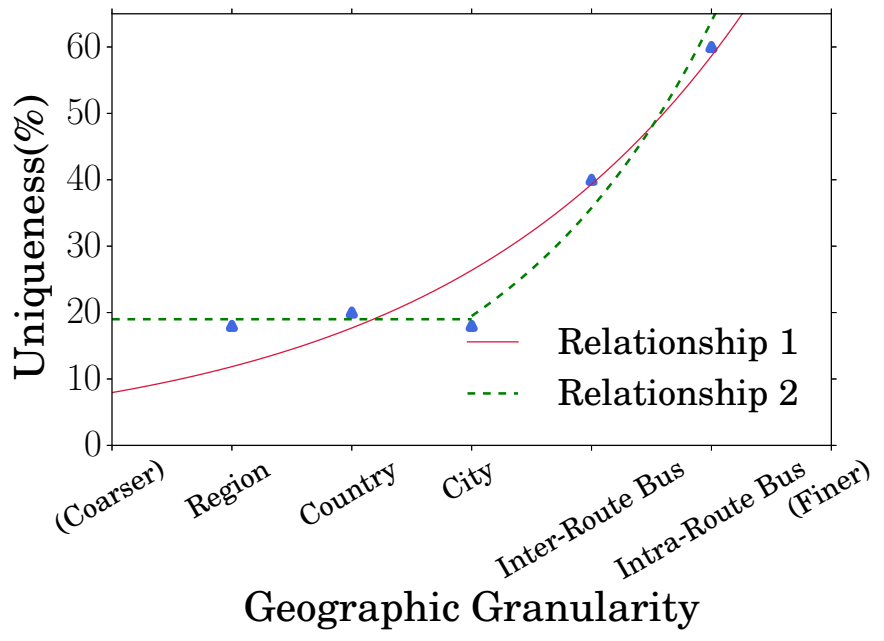

Fig. 3. Possible Geo Correlation vs Geo Granularity relationship.

## 3 uSTASH: DESIGN AND MODEL

We first introduce *uStash* system and then proceed to model the bandwidth costs (for both users and the public transportation company[1]) and the performance of our system with metrics such as hit rate and completion time.

### 3.1 uStash System Design

*uStash* aims to provide an efficient and cheap mobile content distribution with higher *QoE* in public transportation systems. This is achieved by allowing the users to contribute to the distribution of content by pushing any content that they may have consumed (while using the *uStash* service) into a local stash (*US-Server*) via a local network (*US-LAN*) as shown in Fig. 4. US-proxy is a cloud server that could be physically located anywhere, and it is located in the lab for our implementation. US-server is on-board inside buses or trains, while US-LAN is the local WiFi connection between users and US-server. If the majority of the content can be distributed locally by the *US-Server*, the users' and stash provider's cellular bandwidth usage will be reduced and latency minimized thus improving the user *QoE*.

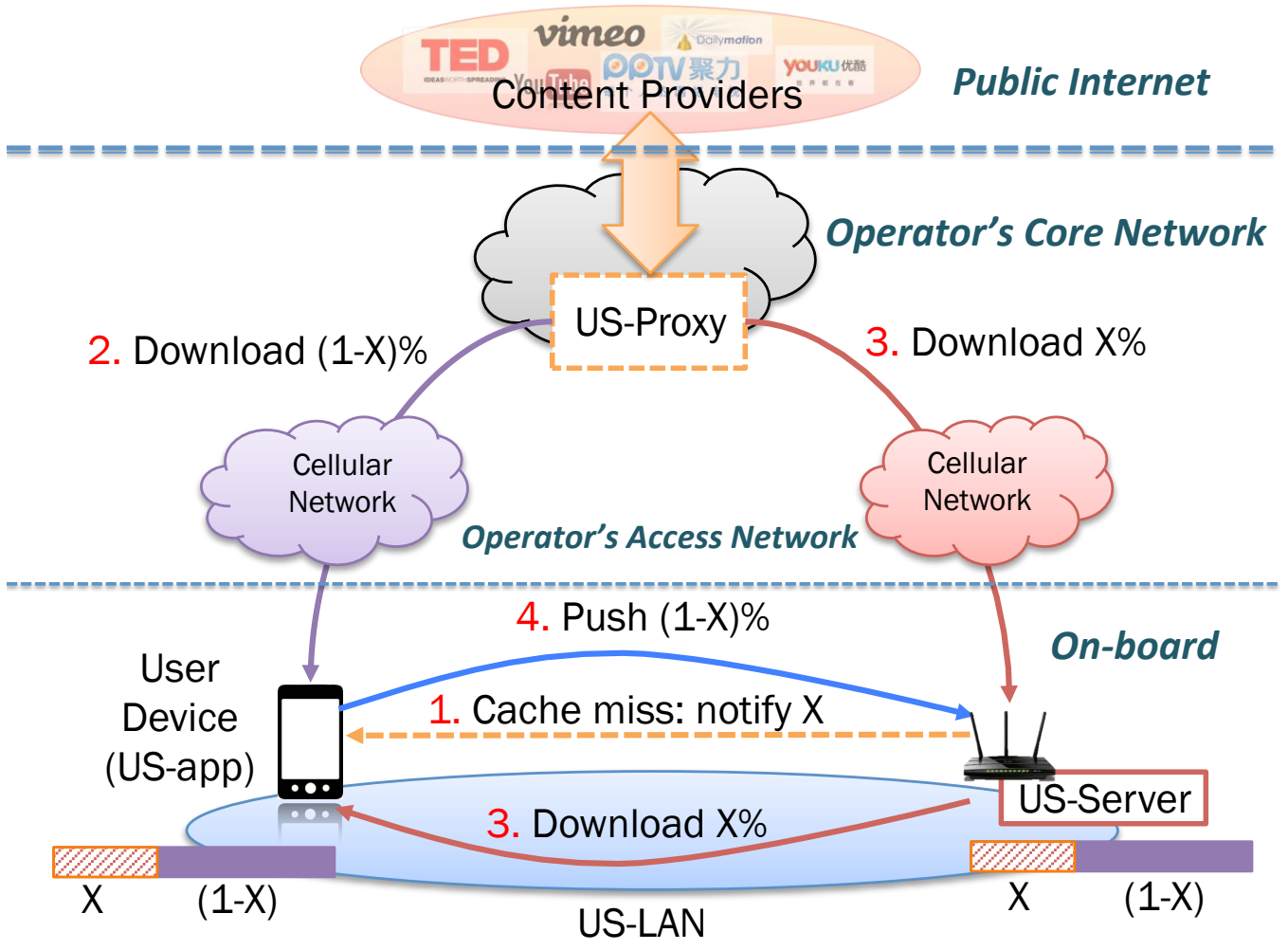

Fig. 4. *uStash* system overview

Typically, users access the *uStash* system via an application (*US-app*) installed on their devices or via an extension to their mobile web-browsers. As they get on the bus and request any content through the *US-app*, the *US-app* automatically switches the user device to *US-LAN*[2] and forwards the request to the *US-Server*. If the content is available in the *US-Server*, it is served to the user via *US-LAN* locally. If not, the content will be cooperatively downloaded by the stash provider and the user as shown in Fig. 4. The portion of content to be downloaded by the stash provider $x : (0 \leq x \leq 100\%)$ is dynamically determined by the *US-Server* which informs *US-app* (step 1). $x\%$ is determined by considering contextual parameters of the network and the requested content such as popularity of the content and quality of the cellular network connection of the stash at the time of request, which is further described in Section 3.4. Since the original content providers (e.g. YouTube) normally do not support content split into sub-parts, all the user

1. In the remainder of this paper, assume that the public transportation company is responsible for the deployment of *uStash*. We then use public transportation company and stash provider interchangeably.

2. The access to the *US-LAN* is managed by *US-app* using standard WiFi authentication without any user interaction.

TABLE 3
Summary of symbols for system model

| Symbol | Description |
|---|---|
| $C$ | set of total available content |
| $M$ | total number of available content $\lvert C\rvert$ |
| $N$ | total number of requests |
| $p_k$ | Zipf distributed popularity of $k^{th}$ ranked content |
| $s$ | exponent of Zipf distributed $p_k$ |
| $\mathbb{E}(Y)$ | expected number of unique content after $N$ requests |
| $s_k$ | Gamma distributed size of content $k$ |
| $k, \theta$ | shape and scale parameters of Gamma distributed $s_k$ |
| $V_{r,k}$ | View time normalized by length of the content $k$ |
| $\lambda_e$ | exponent of the Exponentially distributed $V_{r,k}$ |
| $\phi_b$ | cost per mega-byte for the stash provider |
| $\phi_u$ | cost per mega-byte for the user |
| $t_u$ | time taken to download $y_k$ portion by the user |
| $t_b$ | time taken to download $x_k$ portion by the stash |
| $C_b$ | total cost for the stash |
| $C_u$ | total cost for all users |
| $x_k$ | portion of content $k$ downloaded by the stash provider |
| $y_k$ | portion of content $k$ downloaded by the user |
| $\omega_b$ | cellular network bandwidth for the stash provider |
| $\omega_u$ | cellular network bandwidth for a user |
| $\omega_l$ | local WiFi network bandwidth |
| $T$ | expected completion time of a content request |

traffic is redirected via the *US-Proxy*, which is a server in the cloud, as shown in Fig. 4. *US-Proxy* makes sure original content is split into sub-parts such that it can be aggregated at the user device. The user downloads from *US-Proxy* $(1-x)\%$ of the video using his personal cellular network connection (step 2) while the remaining portion $x\%$ will be downloaded by the stash and delivered to the user via *US-LAN* (step 3). While the user consuming the content, e.g. watching a video, the $(1-x)\%$ of the content will be pushed to the *US-Server* by the *US-app* without disrupting the user as well as without any user intervention (step 4). The proactive push (from users to the *US-Server*) of the $(1-x)$ sub-part of downloaded content to the stash populates the *US-Server* by content that is likely to be popular in *US-Servers* geographical area. Details of *US* implementation and real-world experiment results are presented in Section 5.

The stash provider may advertise a minimum value for $x = x_{min}$ for the purpose of providing guaranteed savings for the users on-board even in the case of a stash miss. For instance, if $x_{min} = 20\%$, all users will save at least 20% of their cellular bandwidth in addition to the *QoE* improvement. Moreover, in the case of a stash miss, *US-Server* can recommend a set of similar content to the requested content which are already stashed in the *US-Server*. Krishnappa et al. [15] shows that viewers are more likely to watch content in the top recommended list and increases stash hit rates by 2 to 5 times. We also believe that *uStash* users would have higher likelihoods of watching locally stashed content as it not only provides better *QoE*, but does incur no downloading costs.

## 3.2 User Incentives

The incentives of using *uStash* is driven by the willingness to save cellular data and improve video *QoE*. Moreover, energy consumption is reduced as shown in Section 5.1 for the reason that WiFi consumes less energy compared to cellular when downloading the same amount of data. Users are also able to use their own device and consume videos across major content providers without the need of interacting with the system. Finally, users are able to consume only the free videos locally stashed in their proximity completely for free, while *uStash* guarantees lower cost and better *QoE* for external videos users request.

## 3.3 System Model

In this section, we consider modeling the costs (due to downloading of both service providers and users) and performance of the *uStash* system by considering the expected completion time of content download. Then we examine the optimal portion of content that should be downloaded by the stash provider to satisfy either of the following three objectives: (i) Minimize expected completion time for each content (ii) Minimize system cost and (iii) Optimize system performance. Later in Section 4, we validate our models using real-life data driven simulations and real-world experiments.

### 3.3.1 Content access pattern

We consider a scenario that a set of commuters in a bus requesting content from a pool of web-content $C$, where total number of available content $M = |C|$. We model cost and performance of *uStash* system after $N$ number of requests have been issued to the system. The requests are modeled using content access patterns observed in large-scale datasets, including content popularity distribution, size distribution and content consumption ratio.

*Content popularity:* It has been observed that mobile content popularity follows Zipf-like distributions [16]. We define the popularity of $k^{th}$ ranked content out of all $M$ available content as $p_k = \frac{1/k^s}{J_{M,s}}$, where $s$ is the parameter characterizing the Zipf distribution and $J_{M,s}$ is the $M^{th}$ generalized harmonic number such that $J_{M,s} = \sum_{n=1}^{n=M}(1/n^s)$.

The probability that content $k$ is requested at least once after $N$ requests is $\big(1-(1-p_k)^N\big)$. We denote $C_N$ as the set of requested content. Therefore, the expected number of unique content $Y = |C_N|$ after $N$ requests can be defined;

$$\mathbb{E}(Y) = \sum_{\forall k \in C} 1-(1-p_k)^N$$

$\mathbb{E}(Y)$ is dependent on the exponent $s$ of the popularity distribution as we assume $M$ is significant for mobile web-content. Once content $k$ is downloaded, the content will be stashed and distributed locally. As a result, $\mathbb{E}(Y)$ would be the number of downloads that the *uStash* system is required to perform using the cellular network resources.

*Content size:* The size distribution of content is considered as Gamma distributed [17] such that the size of content $k$, $s_k \sim \Gamma(\beta, \theta)$. We observe that there is a significant difference in size between non-video and video content $C_v$ in *DS1*, although each category individually follows a Gamma distribution. Therefore, we denote $s_k \sim \Gamma(\beta_v, \theta_v) : k \in C_v$.

*Content view ratio:* It has been shown that more than 80% of videos have less than 0.3 view ratio $V_{r,k}$ (view time normalized by the length of the video) [16]. Moreover, $V_{r,k}$ is closely related to the size of the video, where larger contents are less likely to be fully consumed. This behavior has to be considered in *uStash* as it determines the significance of stashing the full content. We model the viewing ratio as an exponentially distributed variable; $V_{r,k} \sim \exp(\lambda_e)$.

#### 3.3.2 Cost of content access

Network coverage, channel quality and monetary cost per Byte could vary amongst different mobile network operators. Let the download cost per mega-Byte for the stash provider, e.g. bus operator, be $\phi_b$ and $\phi_u$ for the user. Then, the cost of requesting content $k$ for the stash and for the user can be defined as;

$$c_b(k) = \phi_b s_k x_k \quad \text{and} \quad c_u(k) = \phi_u s_k y_k$$

where $x_k$ is the ratio of content downloaded by the stash, $y_k$ is the ratio of content downloaded by the user, where $V_{r,k} \leq x_k + y_k \leq 1$. In the remainder of the analysis, we use $x_k + y_k = V_{r,k}$ without loss of generality as the system does not aim to download more than the requested amount of content. The total cost for the stash ($C_b$) would then be computed by multiplying $c_b(k)$ by the probability of requesting content $k$ at least once during $N$ requests $(1-(1-p_k)^N)$ and summing over all available content;

$$C_b = \sum_{\forall k \in C} \phi_b s_k x_k (1-(1-p_k)^N) \tag{2}$$

Similarly, the total cost for users ($C_u$) can be expressed as;

$$C_u = \sum_{\forall k \in C} \phi_u s_k y_k (1-(1-p_k)^N) \tag{3}$$

#### 3.3.3 Quality of Experience

The users' connections are practically more diverse than the stash service providers' connection due to the availability of multiple network operators. Therefore, we assume the channel bandwidth of users, the available cellular bandwidth of the stash provider and local WiFi network are $\omega_u$, $\omega_b$ and $\omega_l$ on average respectively. In addition, we assume that $\omega_l \geq \omega_b \geq \omega_u$ without loss of generality and as such the local WiFi connection are considered not a capacity bottleneck for the system.

We regard the expected completion time of a download $\mathbb{E}(T)$ as the metric of user *QoE*. In the case of a stash miss, the completion time is the longer time taken by either the user $t_u$ or the bus $t_b$ to finish the assigned portion of content. This completion time can then be expressed as;

$$T_{miss} = \max\{t_b, t_u\} = \max\{s_k(V_{r,k} - x_k)/\omega_u, s_k x_k/\omega_b\}$$

For a stash hit, $T_{hit} = s_k/\omega_l$. In addition, there will be $\sum_{\forall k \in C}(1-(1-p_k)^N)$ the number of expected stash misses as it is the expected unique number of content. Conversely, $\sum_{\forall k \in C}(p_k N - (1-(1-p_k)^N))$ is the number of expected cache hits after $N$ number of content requests have reached the system. Therefore, the expected completion time can be represented as:

$$\mathbb{E}(T) = \frac{1}{N} \sum_{\forall k \in C} \max\left\{ \frac{s_k(V_{r,k} - x_k)}{\omega_u}, \frac{s_k x_k}{\omega_b} \right\} (1-(1-p_k)^N)$$
$$+ \frac{1}{N} \sum_{\forall k \in C} \frac{s_k}{\omega_l}(p_k N - (1-(1-p_k)^N)) \tag{4}$$

### 3.4 System Objectives

Next, we study parameters of the collaborative download while minimizing the expected completion time and service costs from the perspective of user and stash provider. We define three main objectives.

**Objective 1: Minimum expected complete time for each content.** $t_u$ and $t_b$ are dependent on the assigned portion of content $x_k$. Therefore, $x_k$ could be dynamically assigned to achieve the minimum possible completion time $\mathbb{E}(T_{min})$ for each content.

**Proposition 1.** *The expected value of optimal $x_k$, $\mathbb{E}(x_{opt}) = \dfrac{\omega_b}{\lambda_e(\omega_u + \omega_b)}$ such that the completion time for a given content is minimum is: $x_k + y_k = V_{r,k}$, $V_{r,k} \sim \exp(\lambda_e)$.*

*Proof.* The download time for the user is $t_u = s_k y_k/\omega_u$ and for the bus $t_b = s_k x_k/\omega_b$. Since $x_k + y_k = V_{r,k}$, the download time $t_u$ can also be represented using $x_k$ such that $t_u = (V_{r,k} - x_k)s_k/\omega_u$. Again, $T_{miss} = \max\{t_u, t_b\}$, the minimum completion time $\min T_{miss}$ for any content can only be obtained when $t_u = t_b$. Therefore,

$$\frac{(V_{r,k} - x_{k,opt})s_k}{\omega_u} = \frac{x_{k,opt}s_k}{\omega_b}$$
$$x_{k,opt} = \frac{V_{r,k}\omega_b}{(\omega_u + \omega_b)}$$
$$\mathbb{E}(V_{r,k}) = \lambda_e^{-1};$$
$$\mathbb{E}(x_{opt}) = \frac{\omega_b}{\lambda_e(\omega_u + \omega_b)}$$

Now when using $x_k = \mathbb{E}(x_{opt})$; $\forall\, k \in C$ and Equation 4, the minimum expected completion time can be computed when $t_b = s_k \mathbb{E}(x_{opt})/\omega_b$.

$$\mathbb{E}(T_{min}) = \frac{1}{N} \sum_{\forall k \in C} \frac{s_k \mathbb{E}(x_{opt})}{\omega_b}(1-(1-p_k)^N)$$
$$+ \frac{1}{N} \sum_{\forall k \in C} \frac{s_k}{\omega_l}(p_k N - (1-(1-p_k)^N)) \tag{5}$$

The equation above can be further simplified if we assume that the sizes of content are equal and that $x_k$ is same for all content such that $s_k = \mathbb{E}(s_k) = \beta\theta$. The minimum expected completion time for each content would then be:

$$\mathbb{E}(T_{min}) \approx \frac{1}{N}\left( \mathbb{E}(Y)\frac{\mathbb{E}(x_{opt})\beta\theta}{\omega_b} + (N - \mathbb{E}(Y))\frac{\mathbb{E}(s_k)}{\omega_l} \right)$$
$$\mathbb{E}(T_{min}) \approx \frac{\beta\theta}{N}\left( \mathbb{E}(Y)\left( \frac{1}{\lambda_e(\omega_u + \omega_b)} - \frac{1}{\omega_l} \right) + \frac{N}{\omega_l} \right)$$
$$\text{where} \quad \mathbb{E}(Y) = \sum_{k=1}^{M} 1 - (1-p_k)^N$$

**Objective 2: Minimum system cost.** The system cost is defined as the sum of total cost for the stash ($C_b$) and the total cost for all users ($C_u$) after $N$ requests. Then, the systems cost $C_{sys}$ can be defined using Equations 2 and 3 as follows;

$$C_{sys} = C_b + C_u$$
$$C_{sys} = \sum_{\forall k \in C} \phi_b s_k x_k (1-(1-p_k)^N) + \sum_{\forall k \in C} \phi_u s_k y_k (1-(1-p_k)^N), \text{ using } y_k = V_{r,k} - x_k$$
$$C_{sys} = \sum_{\forall k \in C} ((\phi_b - \phi_u) x_k + \phi_u V_{r,k}) \, s_k \left(1-(1-p_k)^N\right)$$

Then, $\mathbb{E}(C_{sys}) = \mathbb{E}(C_b) + \mathbb{E}(C_u)$. Similar to $\mathbb{E}(T_{min})$, if we substitute $s_k, x_k$ and $V_{r,k}$ by their expected values assuming they are constants, the expression of $\mathbb{E}(C_{sys})$ can be approximated as follows;
Assuming $x_k = x_c, s_k = \beta\theta, V_{r,k} = \lambda_e^{-1}; \forall\, k \in C$

$$\mathbb{E}(C_b) \approx \phi_b \beta\theta \mathbb{E}(Y) \cdot x_c \tag{6}$$
$$\mathbb{E}(C_u) \approx -\phi_u \beta\theta \mathbb{E}(Y) \cdot x_c + \phi_u \beta\theta \lambda_e^{-1} \mathbb{E}(Y) \tag{7}$$
$$\mathbb{E}(C_{sys}) \approx (\phi_b - \phi_u)\beta\theta\mathbb{E}(Y) \cdot x_c + \phi_u \beta\theta\lambda_e^{-1}\mathbb{E}(Y)$$

$\mathbb{E}(C_{sys})$ is a linear function of $x_c$ and monotonically decreasing because we assume that the stash provider is able to negotiate with the cellular network provider to pay a cheaper rate than a typical user, i.e. $(\phi_b - \phi_u) < 0$ in general. Therefore, $\min \mathbb{E}(C_{sys})$ would be when $x_c = 1$ such that;

$$\min \mathbb{E}(C_{sys}) = (\phi_b - \phi_u)\beta\theta\mathbb{E}(Y) + \phi_u\beta\theta\lambda_e^{-1}\mathbb{E}(Y)$$

$x_c = 1$ represents the case of free on-board WiFi systems, where all content is downloaded by the service provider. On the other hand, $x_c = 0$ represents the case where no service is provided, and *uStash* simply acts as a "dumb" local stash.

In addition, the stash service provider may advertise a guaranteed amount of bandwidth savings for users purely for marketing purposes, i.e. $0 \le x_c \le x_{max}$. In this case, $\arg\min_{x_c} \mathbb{E}(C_{sys})$ would be either $x_c = 1$ or $x_c = x_{max}$.

**Objective 3: Optimal system performance.** The idea is to optimize the system such that it provides minimal completion time and costs for both users and service providers. We define a metric $H$ combining $\mathbb{E}(T)$, $\mathbb{E}(C_b)$ and $\mathbb{E}(C_u)$ with different ratios to quantify the overall performance as follows;

$$H = \frac{\gamma_t \mathbb{E}(T)}{\max \mathbb{E}(T)} + \frac{\gamma_b \mathbb{E}(C_b)}{\max \mathbb{E}(C_b)} + \frac{\gamma_u \mathbb{E}(C_u)}{\max \mathbb{E}(C_u)} \tag{8}$$

Each individual metric is normalized by its maximum value in the region $0 \le x_c \le 1$ to limit its values between 0 and 1 as it provides the same significance to each metric when adding together. Then, different ratios $(\gamma_t, \gamma_b, \gamma_u)$ provides the flexibility to weight one metric above another, e.g., $\gamma_b > \gamma_u$ makes the cost of the *uStash* provider could be more important than the cost of the user. However, for this evaluation we consider $\gamma_t, \gamma_b, \gamma_u = 1$ giving all three metrics equal significance. Moreover, we consider that view ratio $V_{r,k} = 1$ as it is worst-case scenario for overall performance.

**Proposition 2.** *$\arg\min_{x_c} H = \mathbb{E}(x_{opt}) = \frac{\omega_b}{(\omega_u + \omega_b)}$ when $\gamma_t, \gamma_b, \gamma_u = 1, V_{r,k} = 1$ and $\omega_l > \omega_b > \omega_u$.*

*Proof.* From Equation 4 and assuming $x_k, s_k$ *and* $V_{r,k}$ does not change with $k$ such that $x_k = x_c, s_k = \beta\theta, V_{r,k} = 1; \forall\, k \in C$, $\mathbb{E}(T)$ can be approximately simplified as;

$$\mathbb{E}(T) \approx \frac{\beta\theta}{N} \sum_{\forall k \in C} \max\{\frac{1-x_c}{\omega_u}, \frac{x_c}{\omega_b}\}(1-(1-p_k)^N) + \frac{\beta\theta}{N} \sum_{\forall k \in C} \frac{1}{\omega_l}(p_k N - (1-(1-p_k)^N)$$

$\mathbb{E}(T)$ can be expressed as linear functions of $x_c$ when $t_b < t_u$ where $0 \le x_c \le x_{opt}$ and $t_b \ge t_u$ where $x_{opt} \le x_c \le 1$ as follows;

$$\mathbb{E}(T) \approx \begin{cases} -\frac{\beta\theta\mathbb{E}(Y)x_c}{N\omega_u} + \frac{\beta\theta\mathbb{E}(Y)(\omega_l - \lambda_e\omega_u)}{N\lambda_e\omega_u\omega_l} + \frac{\beta\theta}{\omega_l} & \text{for} \quad t_b \le t_u \\ \frac{\beta\theta\mathbb{E}(Y)x_c}{N\omega_b} + \frac{\beta\theta(N - \mathbb{E}(Y))}{N\omega_l} & \text{for} \quad t_b > t_u \end{cases}$$

When $t_b \le t_u$ $\mathbb{E}(T)$ is a decreasing linear function of $x_c$, $\max \mathbb{E}(T)$ is at $x_c = 0$;

$$\text{For} \quad t_b \le t_u;\ \max \mathbb{E}(T) \approx \frac{\beta\theta\mathbb{E}(Y)}{N\lambda_e\omega_u} - \frac{\beta\theta\mathbb{E}(Y)}{N\omega_l} + \frac{\beta\theta}{\omega_l} \equiv \eta_1$$

When $t_b > t_u$ $\mathbb{E}(T)$ is an increasing linear function of $x_c$, $\max \mathbb{E}(T)$ is at $x_c = 1$;

$$\text{For} \quad t_b > t_u;\ \max \mathbb{E}(T) \approx \frac{\beta\theta\mathbb{E}(Y)}{N\omega_b} - \frac{\beta\theta\mathbb{E}(Y)}{N\omega_l} + \frac{\beta\theta}{\omega_l} \equiv \eta_2$$

$\max \mathbb{E}(T) \approx \max\{\eta_1, \eta_2\}$. As we consider $\lambda_e = 1$, the difference between $\eta_1$ and $\eta_2$ is dependent on $\omega_u$ and $\omega_b$ such that $\eta_1 \sim \frac{1}{\omega_u}$ and $\eta_2 \sim \frac{1}{\omega_b}$. From the fact that $\omega_b > \omega_u$, $\max \mathbb{E}(T) = \eta_1$.

Using Equations 6 and 7;

$$\max \mathbb{E}(C_b) = \phi_b\beta\theta\mathbb{E}(Y)$$
$$\max \mathbb{E}(C_u) = \phi_u\beta\theta\mathbb{E}(Y)\lambda_e^{-1}$$

Then, we take two cases of $t_b \le t_u$ and $t_b > t_u$ separately to find the $\arg\min_{x_c} H$ for each case. For $t_b \le t_u$;

$$H = \frac{\gamma_t\mathbb{E}(T)}{\eta_1} + \frac{\gamma_b\mathbb{E}(C_b)}{\phi_b\beta\theta\mathbb{E}(Y)} + \frac{\gamma_u\mathbb{E}(C_u)}{phi_u\beta\theta\mathbb{E}(Y)\lambda_e^{-1}}$$
$$H = \underbrace{\left(-\frac{\beta\theta\mathbb{E}(Y)\gamma_t}{N\omega_u\eta_1} + \gamma_b - \lambda_e\right)}_{m_1} \cdot x_c + \frac{\gamma_t}{N\omega_u} + \gamma_u$$

$V_{r,k} = 1$ makes the $\lambda_e = 1$ in the above equation. When $\gamma_t, \gamma_b, \gamma_u = 1, \lambda_e = 1$ and $\omega_l > \omega_u$, $m_1$ is negative and thus the minimum value for $H$ would be at the largest $x_c$, which is $x_c = x_{opt}$ due to the $t_b \le t_u$ constraint.
Similarly, now we consider the case for $t_b > t_u$;

$$H = \underbrace{\left(\frac{\beta\theta\mathbb{E}(Y)\gamma_t}{N\omega_b\eta_1} + \gamma_b - \lambda_e\right)}_{m_2} \cdot x_c + \frac{\gamma_t\beta\theta(N - \mathbb{E}(Y))}{N\omega_l\eta_1} + \gamma_u$$

When $\gamma_t, \gamma_b, \gamma_u = 1, \lambda_e = 1$ and $\omega_l > \omega_u$, $m_1$ is positive and thus the minimum value for $H$ would be at the smallest $x_c$, which is $x_c = x_{opt}$ due to the $t_b > t_u$ constraint. Thus, $\forall\, x_c$, $\arg\min_{x_c} H = \mathbb{E}(x_{opt}) = \frac{\omega_b}{(\omega_u + \omega_b)}$. □

Based on Proposition 2, $\min H$ could then be formulated by substituting $x_c = x_{opt}$;

$$\min H = 1 + \frac{1}{N\omega_u} - \frac{\omega_l\omega_b\mathbb{E}(Y)}{(\mathbb{E}(Y)(\omega_l - \omega_u) + N\omega_u)(\omega_u + \omega_b)} \tag{9}$$

In the evaluation section, we compare these analytical results for $\min H$, $\mathbb{E}(T)$, $\mathbb{E}(C_b)$ and $\mathbb{E}(C_u)$ are compared with simulation results with real-world dataset.

# 4 Evaluation

We evaluate the performance of the *uStash* system and analyze benefits to the users and the stash provider under various traffic conditions in different geographical environments. We validate the robustness of the proposed models by comparing results of simulations against the analytical results presented in the previous section. Additionally, we compare the system performance with the following three different existing settings for accessing content in public transportation systems.

*1) Access through users' individual cellular network connections:* In this scenario, users download content through connections to a cellular network (e.g. subscribing to mobile data plans offered by mobile providers). This scenario does not entail any additional cost to the transport company but represents in turn substantial download costs for users and potentially degraded *QoE*, although it is currently the most common mode of accessing content due to its convenience and simplicity.

*2) On-board free/paid WiFi services:* In this setting, the transport company provides comparatively cheap or free access to Internet. This is potentially the best case scenario for users in terms of cost, but not in terms of *QoE* as the wireless back-haul from the vehicle to the cellular network base station is often with limited resources. Further, this is the worst case setting for the transport company in terms of costs.

*3) On-board WiFi networks with a traditional cache:* Enabling traditional content caching on-board reduces the cost for transport companies and improves *QoE* for users only in the case of cache hits.

## 4.1 Synthetic Dataset Generation and Evaluation Setup

*HTTPS* traffic accounts for a significant portion of today's internet traffic (approximately 50% of the total internet traffic according to [18]). However, transparent proxies are not allowed to capture encrypted *HTTPS* traffic without compromising users' privacy. As such, our dataset of on-board *WiFi* users (*DS1*) does not contain *HTTPS* traffic. Consequently a majority of video content requests are missing from *DS1* since video traffic is often encrypted, e.g. *YouTube* traffic. As a result, we create synthetic traces by augmenting *DS1* with additional mobile video content extracted from our mobile on-demand video dataset (*DS2*) in the hope to represent a more realistic commuter data traffic. To preserve the observed spatial and temporal characteristics of the traffic collected from each bus running on specific routes and timetables, the additional video contents are generated to match the observed content request patterns.

More specifically, the four weeks of content requests patterns of *DS1* and *DS2* are aligned in time with an hourly granularity. Then, we randomly draw $N_v(h)$ number of video content from *DS2* out of the total available video content $N_{ds2}(h)$ in the particular hour $h$. $N_v(h)$ is proportional to $M_{ds1}(h)$ the total number of requests generated by a specific bus in *DS1*. By varying the ratio of mixing $R_v(h) = M_{ds1}(h)/N_v(h)$, the concentration of videos in the combined dataset, namely *DS3*, can be controlled to create different workloads for evaluation. In the case of $N_v(h) < M_{ds1}(h)/R_v(h)$, only $N_{ds2}(h)$ number of requests would be made. Thus, *DS3* represents both the observed fine granularity spatio-temporal information and real-world mobile video content consumption pattern. In *DS1*, video content accounts for 1.2% of total requests. Assuming there could be the same amount of *HTTPS* video content ($R_v \sim 83.3\%$), we regard $R_v(h) = 80$ unless otherwise specified. In the later part of the section, we vary $R_v$ to observe the impact of video concentration on the performance results. To preserve the geographical characteristics in *DS2*, we create three different synthetic traces combining *DS1* with each city in *DS3*.

TABLE 4
Summary of analytical model parameters

| Parameter | Value | Parameter | Value |
|---|---|---|---|
| $s$ | 0.716 | $\omega_l$ | 6Mbps |
| $\phi_b$ | 3c/MB | $\omega_u$ | 500kbps |
| $\phi_u$ | 10c/MB | $\omega_b$ | 800kbps |
| $N_{nv}$ | 120,627 | $N_v$ | 1,418 |
| $E[Y]_{nv}$ | 99,576 | $E[Y]_v$ | 1,389 |
| $\theta_{nv}$ | 2102 | $\theta_v$ | 194,061 |
| $k_{nv}$ | 0.0006 | $k_v$ | 0.64 |
| $\lambda_{e,nv}$ | 1 | $\lambda_{e,v}$ | 2.77 |

An event-driven simulator based on *Python* is developed to evaluate the performance of *uStash*. We assume that all requested contents are cacheable. Further, we first set no limitation for the cache size in order to investigate a feasible cache size required in each bus to achieve desired performance. On average, each bus consumes $N_{nv} = 120,627$ non-video contents and $N_v = 1,418$ video contents after four weeks. In case of non-video content, we assume that the whole content needs to be downloaded (i.e., $V_{r,k} = 1$). If the user consumes a portion of a video content, the requested portion of the content is stashed only.

For the analytical evaluation, the simulator uses parameters summarized in Table 4. These parameters are extracted from the *DS3* created with $R_v = 80$. We model the content popularity in *DS3* (using non-linear least squares fitting) as Zipf distributed with the exponent $s = 1.005$. Since there is a big difference in size between video and non-video content, we model the size separately with each category following a Gamma distribution with distinct scale and shape parameters.View ratio ($V_{r,k}$) for video content is modeled as exponentially distributed with parameter $\lambda_e = 2.77$. Network bandwidths for the user, the stash and the local network is considered as $\omega_u = 500\text{kbps}, \omega_b = 800\text{kbps}, \omega_l = 6\text{Mbps}$ respectively as suggested in [19], [20]. The download cost per mega-byte for the stash (corporate/business connections), $\phi_b = 3\text{cent/MB}$ is considered smaller than the cost for individual users $\phi_u = 10\text{cent/MB}$ as per the advertised rates by a mobile operator [21]. However, we vary these parameter values when required to study their impact on the performance metrics under different scenarios.

## 4.2 System Performance Analysis

### 4.2.1 Completion Time Analysis

In Fig. 5a, we show $\mathbb{E}(T)$ values in 4 selected buses on different routes. We observe that although the average completion time does vary across different buses due to the content size difference, $\mathbb{E}(T)$ changes according to the bus download ratio $x_k$. Hence, in Fig. 5b, we show the normalized completion time ($\mathbb{E}(T)/\max\mathbb{E}(T)$) as function of $x_k$ as per our analysis in Section 3 and as per the simulations results using *DS3*. The analytical $\min\mathbb{E}(T)$ is at $\mathbb{E}(x_{opt}) = \dfrac{\omega_b}{\lambda_e(\omega_u + \omega_b)} = 0.62$ using the parameters in Table 4. This shows that by carefully selecting the optimum download ratio, the expected completion time can be reduced by as much as 60% compared to the use of commuters individual cellular network connections (i.e. when $x_k = 0$) and by 25% compared to on-board free WiFi services (i.e. when $x_k = 1$). Then, we compare the analytical $\mathbb{E}(T)$ to the simulated $\mathbb{E}(T)$ for three different cities. The minor observed difference is mainly due to the abstraction of content size, view ratio and popularity distribution are slightly different to the actual dataset. However, we clearly observe that the analytical abstraction reasonably models the expected completion time, and the simulated results validate the correctness of our system model. Note that interestingly $\mathbb{E}(T)$ does not vary significantly across the different cities, which suggests *uStash* maintains good performance even in different geographical region.

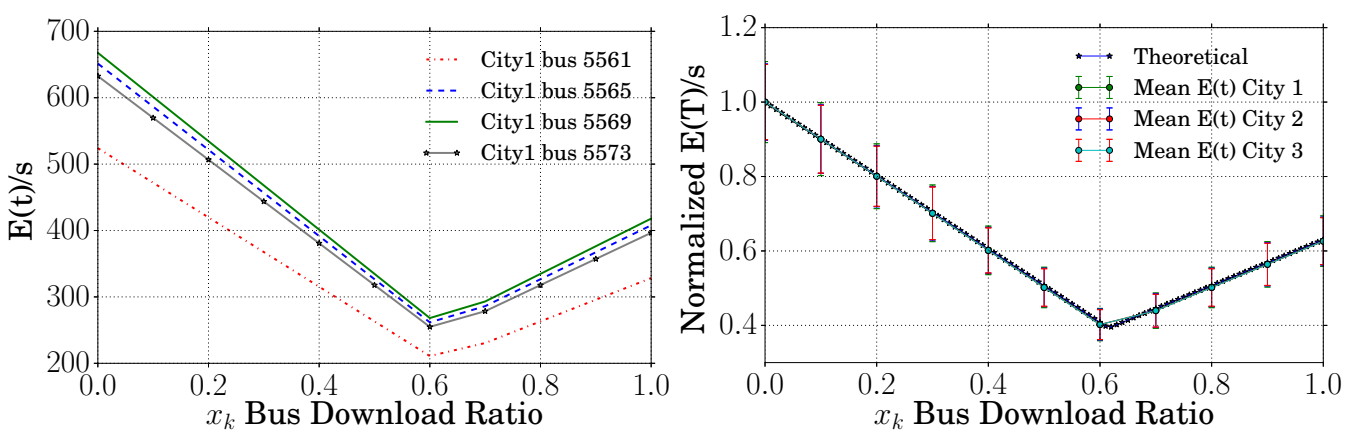

(a) (b)

Fig. 5. Expected $\mathbb{E}(T)$ in different cities and buses.

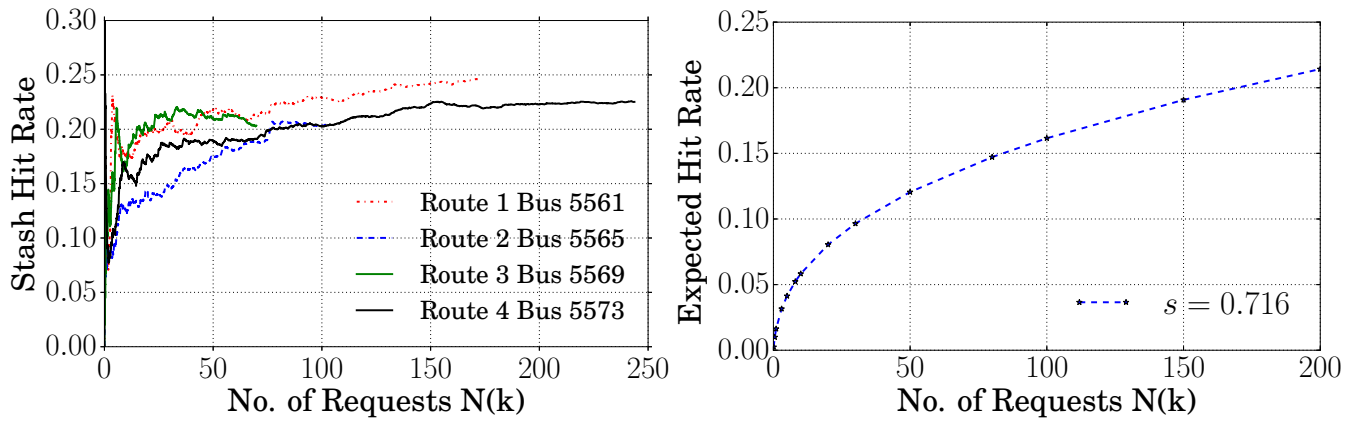

(a) Simulated HR (b) Analytical HR

Fig. 6. Simulated and Analytical Stash Hit Rates.

#### 4.2.2 Stash Hit Rate Analysis

Fig. 6a shows the simulated stash hit rate of a bus with increasing number of total requested content $N$ ($N$ is proportional to time). The results show that irrespective of the geographical route, each bus reaches approximately 20% stash hit rate after a short period of time. The mean stash hit rate for all 22 buses is 21% after four weeks. Fig. 6b illustrates the analytical expected hit rate for a bus. In Section 3, we show that after $N$ requests, we expect $\mathbb{E}(Y)$ number of cache misses and $(N - \mathbb{E}(Y))$ number of cache hits. Therefore, the expected cache hit rate after $N$ requests would be $(N - \mathbb{E}(Y))/N$. Recalling that $\mathbb{E}(Y) = \sum_{k=1}^{M} 1 - (1 - p_k)^N$, we observe that the stash hit rate is considerably dependent on the shape of popularity distribution $s$ in Fig. 6b. The analytical expected hit rate for $s = 0.716$ (extracted from dateset) is inline with the simulated result($\sim 20\%$), although there are slight variations in individual buses. Specifically, in practice, stash hit rate quickly reach the steady state in individual bus comparing to the analytical averaged result. This again, indicates the strong spatio-temporal correlation of content access at the edge of mobile network.

#### 4.2.3 Impact of the size of the stash

The overall traffic during 28 days for the 22 monitored buses is 878GB, resulting in an average 40GB data usage per bus per month. While it is not a small amount of traffic in mobile network, we believe the stash size could be easily implemented to be much larger with a very reasonable cost in each bus. i.e. 128GB SD cards or Terabytes of hard drives. Therefore, we consider that stash size would not be a limitation for *uStash*, although the performance could benefit from cache replacement policies in case stash becomes full. In practice, pre-fetching stash content and periodically swapping popular content could raise a higher demand for storage. However, content caching performances with restricted cache size under various caching replacement policies has also been discussed in a number of works [22], and this paper will focus on the analysis of system performance instead.

### 4.3 Impact of mobile edge popularity distribution

We then show the change of system performance under different content popularity distributions at the mobile edge. More specifically, we vary the shape parameter s of zipf content popularity distribution while keeping other parameters the same. The performance both in the expected hit rate and completion time $\mathbb{E}(T)$ are presented. In Fig. 9, it is shown that the expected hit rate increases as shape parameter s and number of requests N(k) increase. The larger shape parameter s is, the more skewed edge popularity distribution is, which is associated with a higher probability of requesting the same content, leading to higher cache hit rate. Moreover, expected completion time does not vary much for the range $s < 0.8$ before dropping dramatically between $s = 0.8$ and $s = 2$. This is also likely to be caused by a higher portion of content being served locally. However, the design of *uStash* allows an improvement in system performance regardless of the edge content popularity distribution.

### 4.4 System Cost Analysis

As stated earlier, the system cost is quantified by the amount of bandwidth usage by the system. Fig. 7 depicts the stash bandwidth saving by having an on-board stash for Route 1 buses in three different cities. The results show that there is 10-12GB of saving per month for the bus #5562 and between 3 to 6GB of savings for other buses. Note that the results do not change significantly for different cities due to the existence of spatial and temporal correlation of content access as observed in Section 2. The value of bandwidth saving in each bus is, again, dependent on the content request rate, which is approximately 182 req/hr in our collected dataset. The bandwidth saving, therefore, will have a much higher potential in more crowded transport network, i.e. NYC and Tokyo. In Fig. 7d, we show the bandwidth saving when *uStash* is enabled on-board compared to a non-caching solution. On average each bus could save approximately 9.4% of total bandwidth. However, depending on the location it could increase as much as 29% which shows the huge potential benefits for the stash provider.

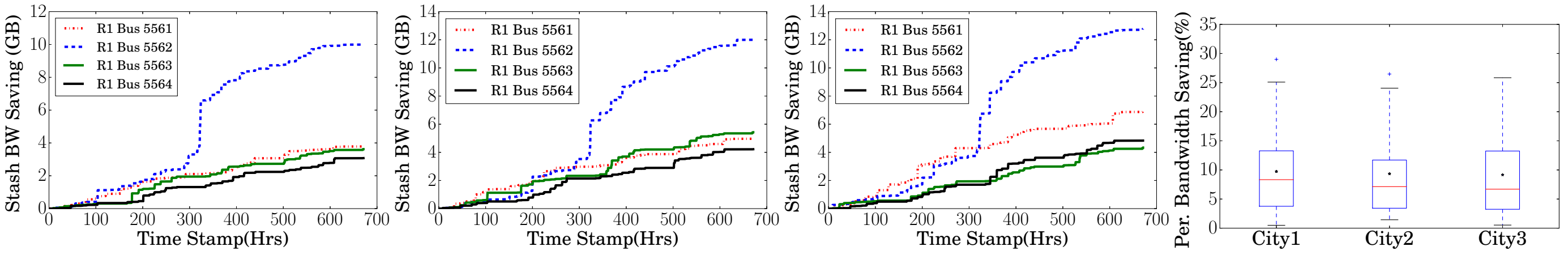

(a) City 1 (b) City 2 (c) City 3 (d) Per. of BW Saving

Fig. 7. Route 1 with City 1-3, Stash Bandwidth saving; Per. of Bandwidth Saving across buses.

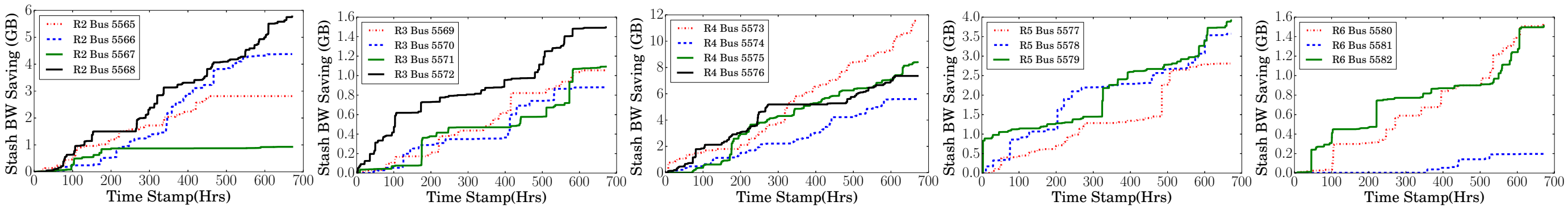

(a) Route 2 (b) Route 3 (c) Route 4 (d) Route 5 (e) Route 6

Fig. 8. Route1 to Route 6 with City 1: Stash Bandwidth saving.

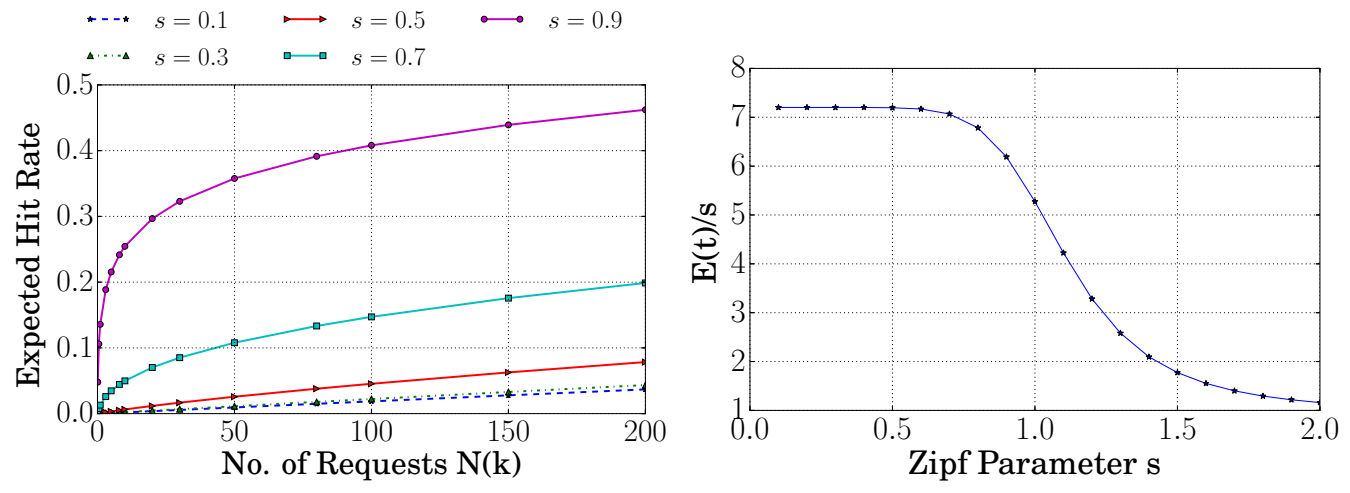

Fig. 9. Impact of edge content popularity distribution

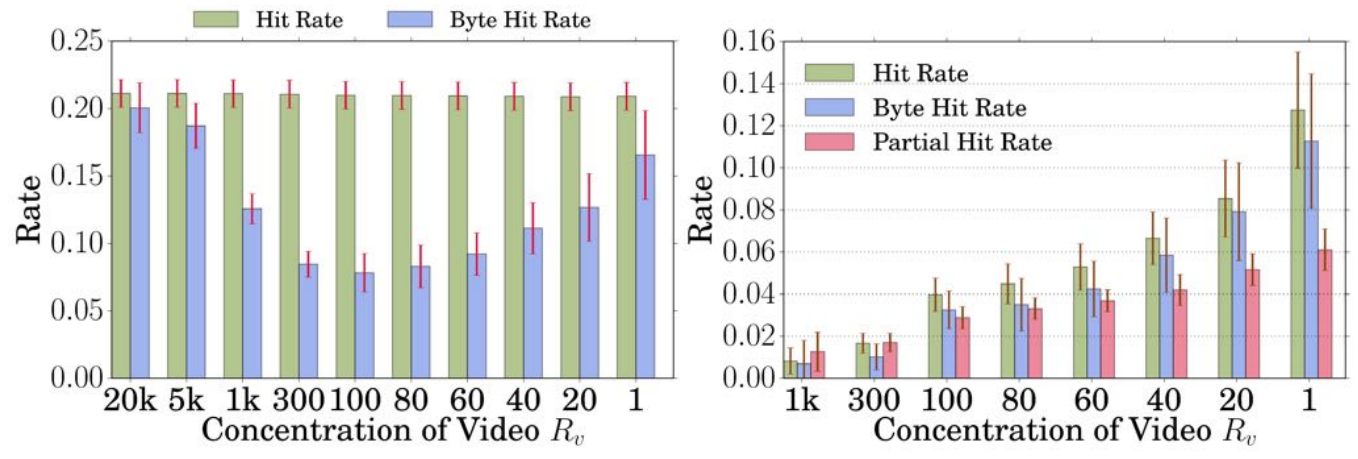

(a) All contents (b) Video Only

Fig. 10. The impact of mobile video concentration.

In Fig. 8, we compare stash bandwidth saving for Route 3 to Route 6 in the same city, which are significantly different to each other. This shows that the spatial locality of content requests at the bus or route level is much more significant than at the city level. This is explained by the fact that cities are large enough to hold a mix of users with different very broad interest, while the bus is smaller and hold users with similar interests.

## 4.5 Impact of concentration of videos

By varying the ratio of mixing $R_v(h)$, we control the amount of video content and generate different synthetic datasets. We show in Fig. 10a the *uStash* performance with regards to stash hit rate and stash byte hit rate. It could be seen that as stash hit rate does not change with the $R_v$. This is due to the fact that the hit rate is dominated by non-video content (video content represents only ~1.2% of all requests). On the other hand, although in smaller number of requests, video content accounts for a higher amount of traffic. As a result, the byte hit rate gradually drops as the $R_v(h)$ decreases before rising up again after $R_v(h) = 100$. The increase of byte hit rate as video concentration drops after threshold could be explained by the fact that byte hit rate is again dominated by non-video content, where non-video content byte hit rate is approximately 20%.

When a user requests a portion of content which is larger than the portion stashed locally, we consider it as a partial hit. For example, previously a user consumes first 20% of a video content, and later another user decides to consume more than the 20% stashed video. Then, the stashed amount is delivered locally while the remaining portion is collaboratively downloaded by the user and the bus similarly to a stash miss. Fig. 10b illustrates the partial hit rate for video content as function of $R_v(h)$. The partial hit rate is significant when there are larger amount of videos, e.g., when $R_v(h) = 1$, the hit rate for videos is approximately 12%, while the partial hit rate $Hit_p$ is half as much (6%). As the video concentration decreases ($R_v(h)$ increases), full hit rate decreases faster than partial hit rate. Note that the partial hit rate becomes larger than the full hit rate for lower concentration of videos (for $R_v(h) > 200$).

## 4.6 System Objective Analysis

In Fig. 11, we compare the expected completion time and system cost of different candidate systems. We observe in Fig. 11a that user direct download would have the worst system performance, whilst *uStash* improves the performance further by approximately 10% comparing to cache enabled WiFi hotspots. In addition in Fig. 11b we compare the system costs with $x_k = \mathbb{E}(x_{opt}) = 0.62$ for *uStash*. On-board WiFi and cacheable WiFi have higher cost to transport company (stash provider) than uStash, and uStash costs higher overall under the assumption that transport company could negotiate a wholesale data cost much lower than users. Although *uStash* is not the most cost efficient system wise, it considerably saves costs for both transport service providers (15%), and for users (65%), when comparing to directly using their data plans. Users could choose to consume stashed content only for free, thus, *US* guarantees users are better off cost-wise compared to no on-board content distribution system.

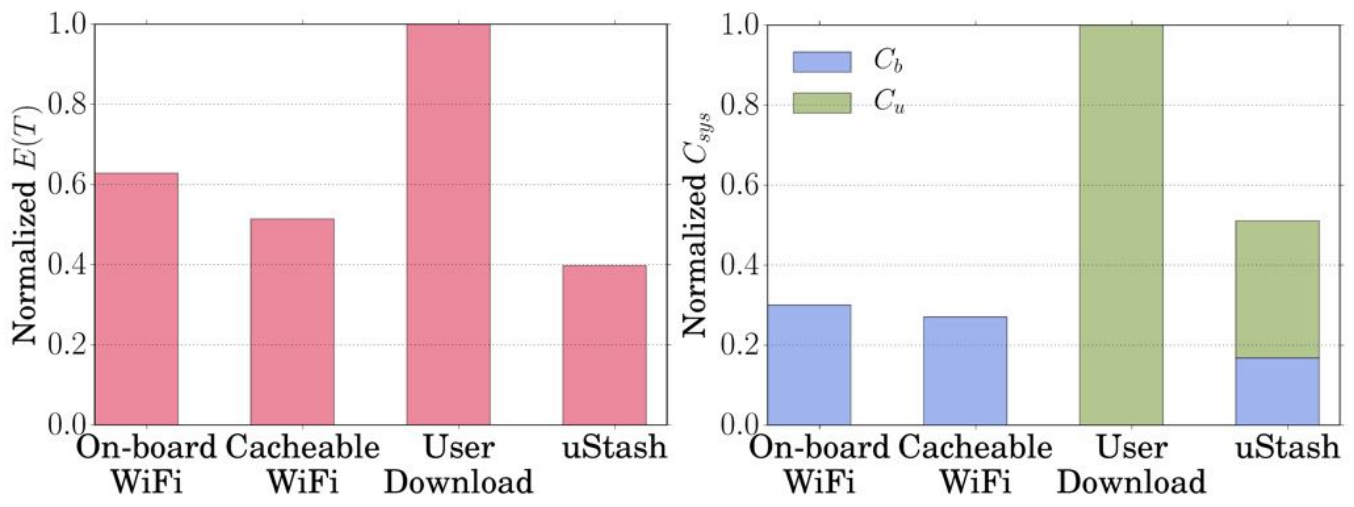

(a) $\mathbb{E}(T)$ (b) $C_{sys}$

Fig. 11. System $\mathbb{E}(T)$ & $C_{sys}$ Comparison.

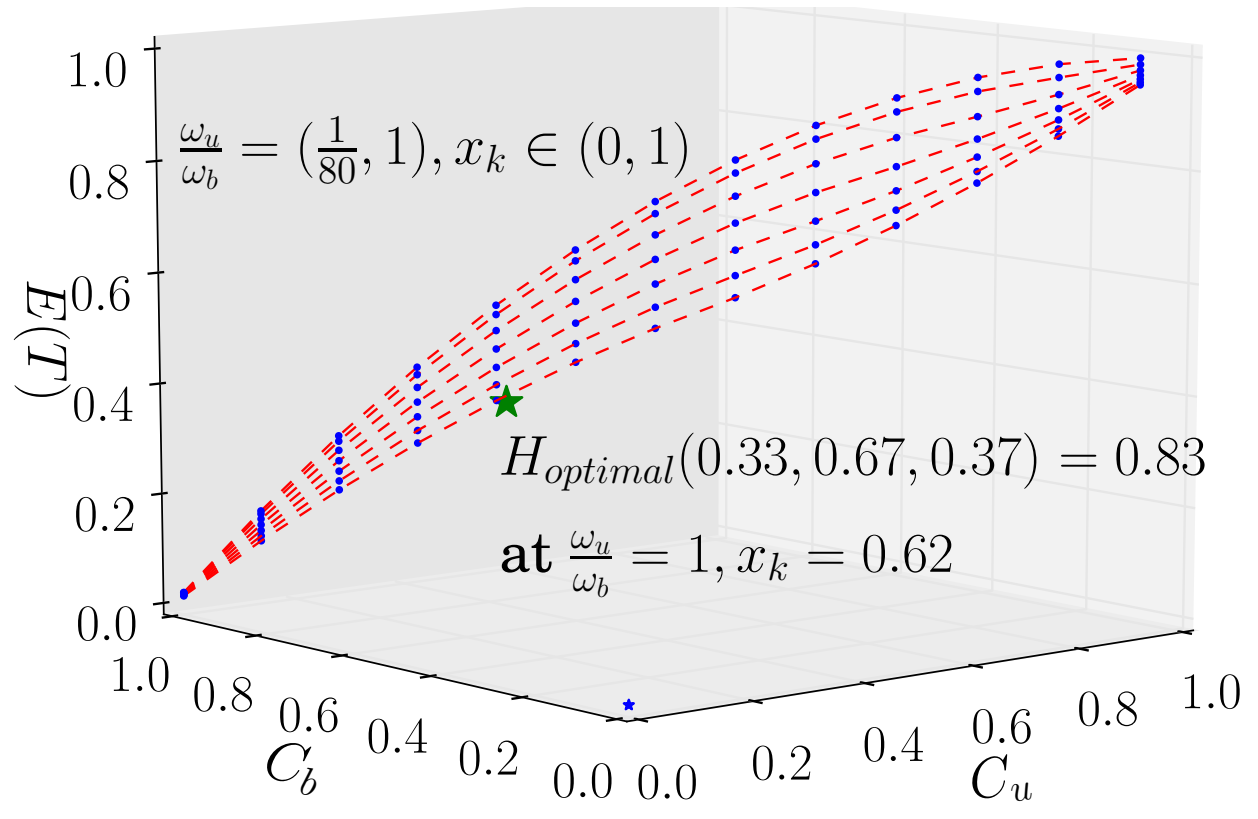

Fig. 12. Modeling the system objective.

We analyze the overall performance in regards to the user cost $\mathbb{E}(C_u)$, the service provider cost $\mathbb{E}(C_b)$ and the system performance metric $\mathbb{E}(T)$ as per Section 3. Fig. 12 shows the relationship between the above three metrics when we vary the service provider download ratio ($x_k$), the ratio of bandwidth between the user and the stash ($\omega_u/\omega_b$). The overall optimum system performance ($H_{optimum}$) is defined as the point with minimum distance to the origin (0,0,0), if we give equal significance to all three metrics i.e. $\gamma_t, \gamma_b, \gamma_u = 1$ as per the Equation 9. Hence, all three performance metrics are normalized in a range between 0 and 1. For instance, the bandwidth ratio of user to the stash ($\omega_u/\omega_b$) is varied from $1/80$ to 1 and $x_k$ is varied from 0 to 1. Based on the figure, the closest point on the surface to the origin is (0.33,0.67,0.37), which represents a distance of 0.83. This $H_{optimum}$ value matches with the $\min H$ calculated from Equation 9 using the above parameters. Additionally, the point furthest away from the origin is (1,0,1) with a distance of 1.41. This is the worst case scenario for system performance representing when internet access is only through users' individual cellular network connections. Therefore, the system objective model can be used to understand the performance of the *uStash* system in different network conditions and user environments.

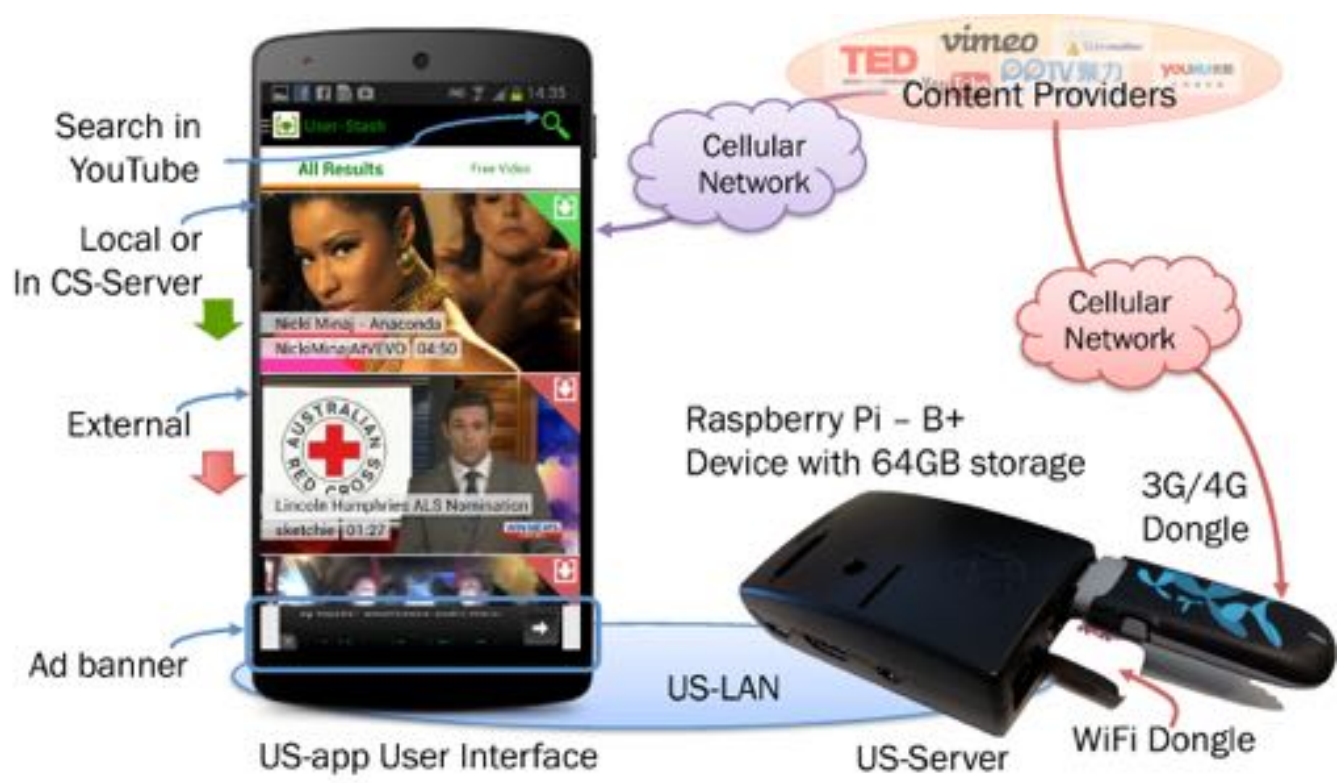

Fig. 13. User interface of US-app and US-Sever.

## 5 Prototype Implementation & Experiment

We implemented *US-app* as an Android app and *US-server* as a *Linux* application on a *Raspberry-Pi*. The low-cost setup could be powered by vehicle's power supply or a mobile power bank. It's a scalable solution for improving mobile user *QoE* in public transport, and in general mobile edge. Although the system concepts are valid for any popular content type, in our implementation we focus on the distribution of video content. The interface of the *US-app* is shown in Fig. 13. When the *US-app* is launched, it requests the unique IDs (URLs) of the most popular videos from the video service providers. The current prototype version is linked with YouTube[3], Dailymotion[4] and local news[5] content via their APIs. The app also allows searching for a particular video or a set of related videos for a particular keyword. Once the relevant video IDs are received, the networking interface switches to the *US-LAN*, when available, to obtain the list of videos that are stashed in the *US-server*. Depending on the user request, the *US-app* displays results, with an indication of whether the content can be obtained from the *US-server* (green arrow) or needs to be downloaded via the cellular network (red arrow). If content can be obtained locally, the *US-app* fetches the content from the *US-server* via the *US-LAN*. Local advertisement distribution mechanisms are also developed for future use. Engineering and hardware implementation of *uStash* could potentially be further optimized, but would not be covered in details in this paper.

### 5.1 "In the wild" Experiment

We conducted an experiment to validate the system with real-world mobile network connectivity and user device Samsung Galaxy S4. Firstly, we show the instantaneous power consumption of user device using *uStash* in a lab setting using fine-grained measurement setup shown in Figure 14. With this setup, the phone is powered by an external DC power supply ($V_p = 5.2V$), and connected in series with a 0.5 Ω shunt resistor $R_s$. Then a National Instrument NI-USB 6008 is used to sample the voltage drop $V_s$, across the shunt resistor at 1KHZ and log the dog onto a computer. To ensure that the power consumption was only

3. https://developers.google.com/youtube/v3/
4. http://www.dailymotion.com/doc/api/graph-api.html
5. http://www.abc.net.au/news/

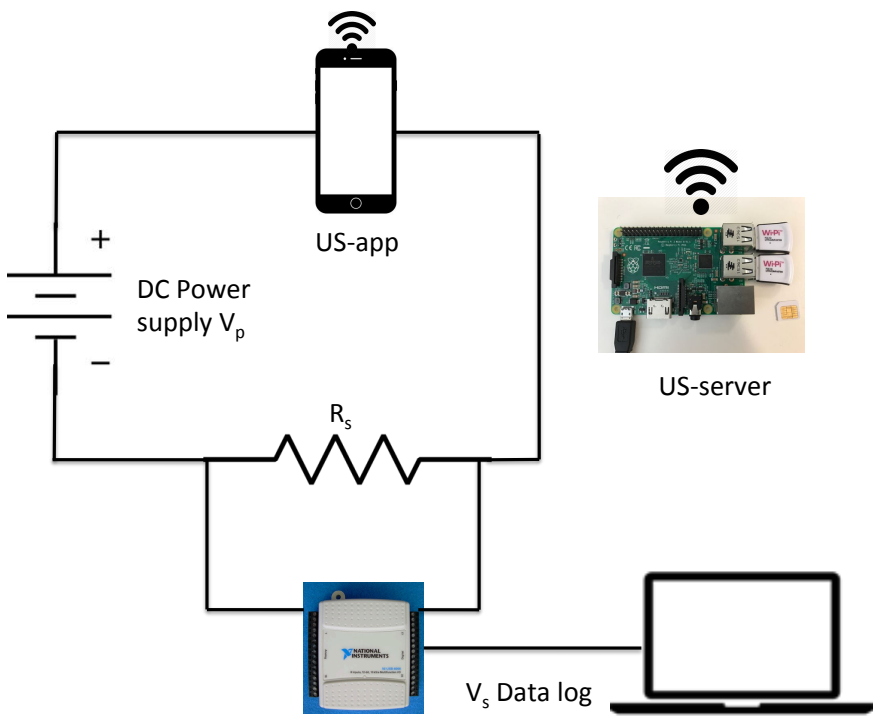

Fig. 14. Power measurement setup

due to the usage of *uStash*, all other processes on the smart phone were terminated via Android developer options. The screen brightness is also set to a fixed level to eliminate any change during the experiment. We then use the standard power equation, $P_{phone} = P_{overall} - P_{shunt} = V_p \times \frac{V_s}{R_s} - \frac{V_s^2}{R_s}$ to calculate the power consumption at each data point.

The instantaneous power consumption of user device with and without using *uStash* are logged and plotted in Figure 15. Moreover, events are annotated onto the consumption traces, where Event 1 to Event 4 are the major events logged in the mobile phone. User device will first check with local US-server whether the requested video is available locally. In our experiment, we show the case of stash miss. Event1:BeginClick is trigger when user starts watching a video by clicking on the video thumbnail. The first $x\%$ if video would be fetched from user's cellular link, and at the same time, US-server will download the remain (1-$x\%$). Event2 is logged when US-app begins downloading first $x\%$ of the video, and Event3 marks the end of downloading first $x\%$ chunks assigned to user device. US-app will then switch to US-LAN to receive the remain (1-$x\%$) chunks (Event4) until user finish watching the whole video or dropping the video by an "end click". It could also be seen that "user download" (case b) consumes higher power for phrase between Event1 and Event3 (red) with a peak power consumption reaching 6W. This is due to a larger amount of data (100%) being downloaded using cellular 3G/4G interface. Furthermore, (b) has a smoother power consumption for the phase upon the end of Event3 (green) compared to (a). The reason is that once finishing downloading the video, user device is playing the video that is locally cached in the user phone as opposed to streaming from US-server in case (a).

Then, an "in the wild" experiment in small-scale with one US-server and one active user is conducted. Data is collected in multiple buses among various routes in Sydney, and both cellular connections of the user device and US-server are provided by a major mobile ISP. The user continuously watches videos during the period of bus trips, and US-app logs event times and video meta-data of all videos watched. The 43 consumed video logs are further analyzed and used to present the results. Software power profiler *BatteryManager* is used to monitor the battery usage. *uStash* is shown to consume slightly lower power (10%) in Fig. 16a due to the fact that downloading or streaming via *WiFi* consumes less power compared to using cellular

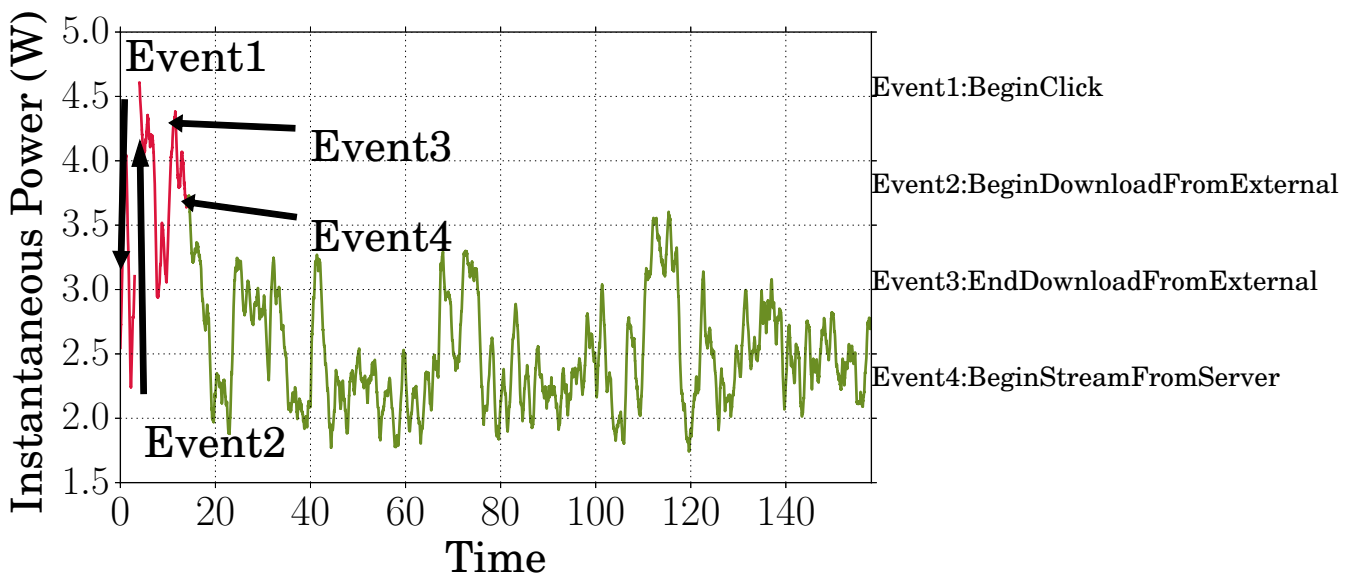

(a) uStash "stash miss"

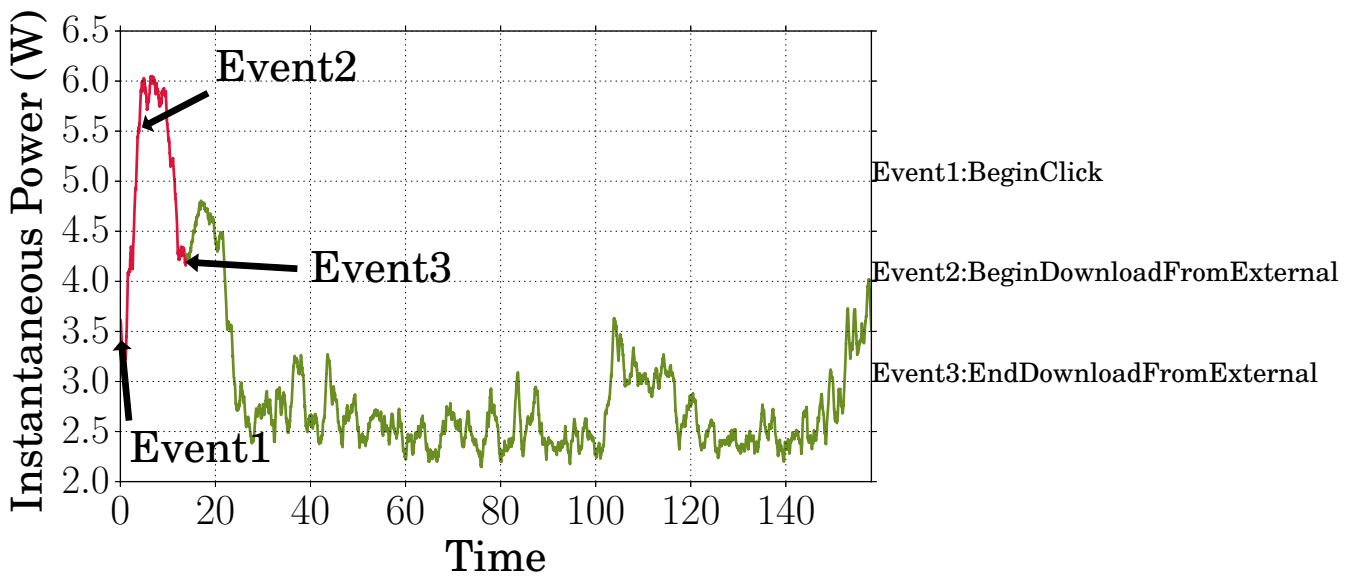

(b) User Download

Fig. 15. Instantaneous Power Consumption

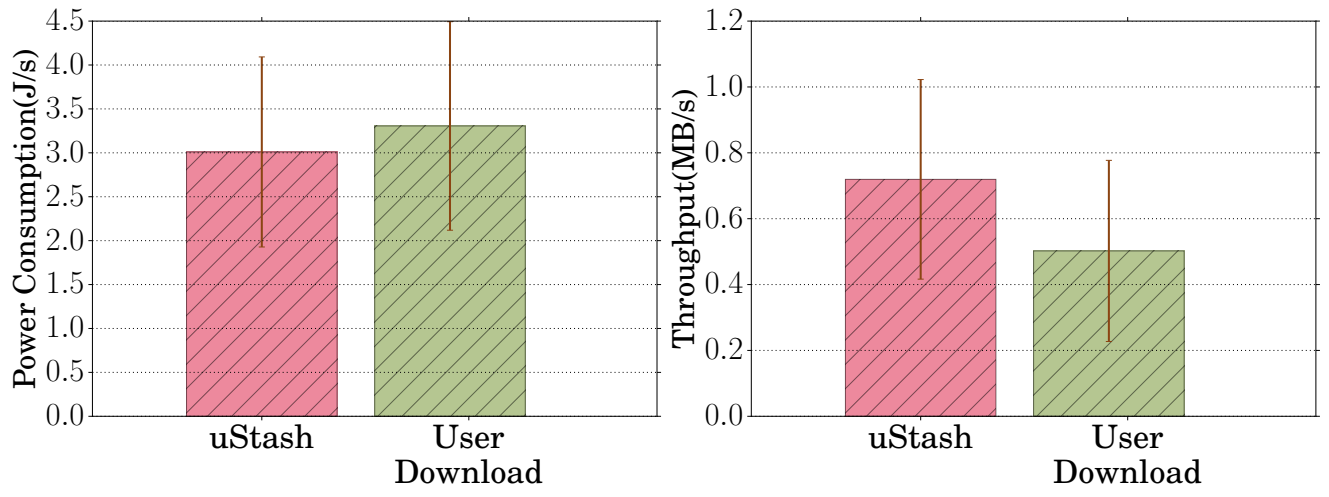

(a) Power consumption (b) System Throughput

Fig. 16. "In the wild" Power consumption and Throughput

network, i.e. 3G and 4G.

Apart from energy consumption, lowered cost and improved *QoE* are other incentives for commuters to use *uStash*. In our experiment, user data cost is reduced by approximately 50%, in the meantime, we present user *QoE* through measurable throughput. As expected, cellular network throughput varies widely depends on location. Majority of the time, throughput is lower than 1MB/s even when most measurement is done in urban areas. We compare the average throughput of *uStash* perceived by user to the scenario when users download the whole video by their own device in Fig. 16b. *uStash* average throughput is calculated by the weighted average of users' cellular throughput and WiFi throughput between user device and US-server. It could be seen that *uStash* has an overall system average throughput of approximately 0.72MB/s, which is 43% higher than the throughput of users' cellular connection alone. Although mobile network performance has significantly improved and is continuously improving, *uStash* is and will be beneficial to mobile user *QoE* due to a). high variance of cellular network performance and b). higher resolution videos/larger size content.

We are planning to distribute *uStash* system in pilot projects with transportation companies and on-board Internet providers, hence experimentally collect large-scale real-world user statistics to validate the system design and performance.

## 6 Related Work

Mobile Internet access and content distribution with limited network connectivity have been studied extensively. The related efforts to our work can be broadly categorized into three groups, namely *opportunistic/cooperative networking*, *caching*, and *pre-fetching*.

*Opportunistic/Cooperative networking* has been investigated by numerous groups, in particular within the ad-hoc networking community. Opportunistic caching [23] suggests replicating content on other user devices when user devices come in contact so that all devices can act as caches for other device as users move. Mashhadi and Lee et al. [24], [25] extend this concept with *WiFi* access points acting as caches which store and push the information from users when they connect. These schemes essentially exploit user mobility to disseminate content. However, individual users are required to store large amount of content and required to provide access to their personal devices to strangers with the associated privacy and security risks. Other concerns include the implications of using the device resources, especially the battery. *uStash* does not require users to store content for others, nor provide access to their devices to strangers. Moreover, the energy usage is minimized as part of the content is downloaded via a local high-speed network, and it is commonly known that using *WiFi* saves energy comparing to cellular [26]. There are also works done in the field of cooperative networking [27], [28]. Padmanabhan et al. [27] proposed cooperative networking to solve the congestion issue of popular servers due to streaming of media contents. *uStash* uses a similar approach, however aims at mitigating the limited bandwidth of last wireless hop. Ribeiro et al [28], regarded wireless user cooperation as a form of multipath. In *uStash* , we operate in the traditional client-server framework and the cooperation is between user and service provider for each specific content.

The research work of mobile caching basically fall into two groups, caching in cellular networks and device caching using social behavior of users. The latter shares the *cooperative* concept with *uStash*. Erman et al. [10] provide a cost-benefit trade-off model to investigate the caching benefits at different levels of a cellular network, e.g. base station nodes and different points within the core network. However, this scheme does not address the congestion on the last hop that needs to be shared by multiple users, which a key objective for *uStash*. Other variants of caching effort, such as forward caching [29] do not address this issue either, since the requested content needs to traverse the last hop at least once per request [2].

*Pre-fetching* schemes (i.e. [2], [3], [30]) attempt to predict the content that users are likely to use in the future, download this content in advance (pre-fetch), and store them locally or in a location closer to the potential content consumers. This approach however requires the ability to predict in advance both the information consumption patterns and the network availability to get the maximum benefit. Firstly, although commuters may have predictable patterns, it is difficult to predict their consumption patterns accurately. Secondly, since there is no content sharing, the proposed schemes do not ultimately reduce the overall network usage. Finally, prediction raises a number of privacy concerns, which have not been fully addressed. In contrast, *uStash* does not rely on any prediction and dynamically adapts to the content demand. As it facilitates sharing, it also reduces the overall network usage.

## 7 Concluding Remarks

In this paper, we first validated our assertion of spatio-temporal correlation of content access on public transport systems using real-world content access patterns of commuters. Then we focused on the challenging problem of providing better *QoE* and cheap Internet access for high density of users in a moving vehicle. We designed, modeled, evaluated and implemented a novel system, called *uStash* that leverages of the spatio-temporal locality and the inherent path diversity provided by the public transport commuters personal mobile connections. *uStash* is designed and optimized both for the routes and scenarios with strong spatio-temporal user correlation and routes with loose user correlations. In contrast to previous attempts, *uStash* provides a clear incentive for user to contribute to the system as it lowers their mobile data consumption, and overcomes the security issues associated with providing strangers access to users personal devices.

We evaluated *uStash*'s performance both analytically as well as through trace driven simulations. We then showed that it is possible to reduce the expected completion time by approximately 60% compared to downloading content using their own cellular connection. Finally, we also implemented the system using *Raspberry-Pi* devices as the stash and Android mobile devices as a *uStash* mobile app to show the practicality of the system.

## Acknowledgment

This research was supported/partially supported by Data61, CSIRO, UNSW and Australian Government Research Training Program Scholarship. We thank our colleagues who provided insights and expertises that greatly assisted the research.

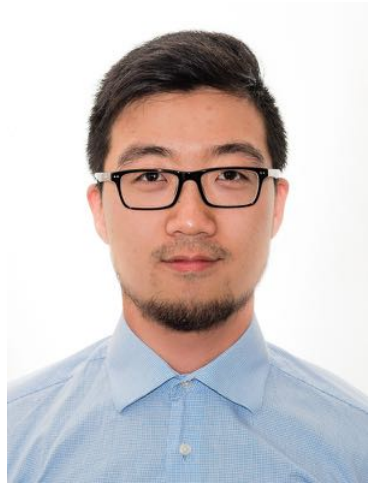

**Fang-Zhou Jiang** is a Ph.D student in school of Electrical Engineering and Telecommunication from UNSW, and an enhanced Ph.D student at Cyber-Physical System Program in Data61, CSIRO. He received the B.E. with First Class Honours in Telecommunications from the University of Sydney, Australia, in 2014. Since 2013, he worked at Network Research Group in NICTA as a summer scholarship student, and later became a research intern. He also spent 6 month in 2016 working as research intern in NII, Japan. His current research interests include Mobile Content Distribution, Mobile Computing and Mobile System QoS/QoE optimization.

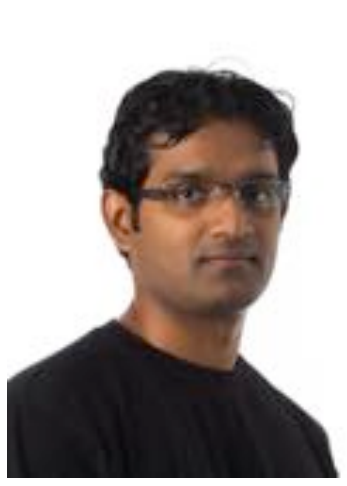

**Kanchana Thilakarathna** is a Researcher at the Mobile Systems Research Group at Data61. He received his Ph.D. in Electrical Engineering and Telecommunications from the University of New South Wales and a B.Sc. degree with First Class Honours from the University of Moratuwa, Sri Lanka. He has more than three years of industry experience as a Mobile Radio Network Optimization Engineer, before pursuing postgraduate studies. Kanchana also holds Conjoint Lecturer appointment at UNSW. His current research interests are in developing novel mechanisms for efficiently delivering mobile data while preserving user privacy, and security when using online services.

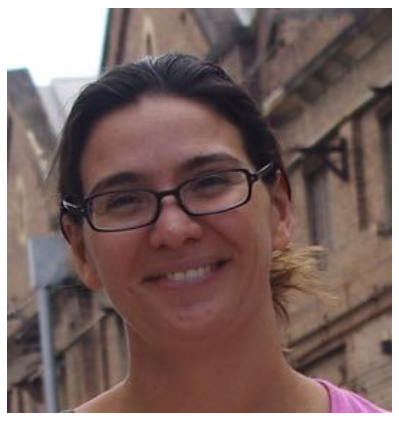

**Sirine Mrabet** is a Research Engineer at the Mobile Systems Research Group at Data61. She joined Data61 in February 2014 and is responsible for the design and the development of mobile networking products. Prior to joining Data61, Mrs Mrabet worked as a project manager for three years at KLS-Logistic Systems, a company dedicated to warehouse management, transport management and optimization. She was responsible of the design, development and delivery of final products to customers including Tag Heuer Swiss (Switzerland), ENTREMONT (France) and CHUs of Lyon, and Saint Etiennne.

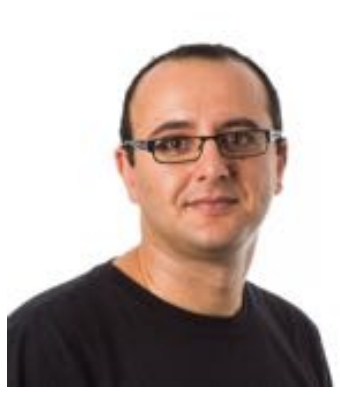

**Mohamed Ali Kaafar** is a research leader at the Mobile Systems Research Group at Data61. He leads the research and development activities in Network measurement, modelling and performance evaluation as well as security, privacy and CyberCrime prevention with a focus on data-driven approaches. He holds the position of visiting professor of the Chinese Academy of Science (CAS). He was previously a researcher at the Privatics team at INRIA in France, and at the university of Liege. He is the main investigator and responsible of several European and Asia-Pacific research projects. Dr. Kaafar obtained an Engineering degree, an M.S from University of Manouba and a Ph.D. in Computer Science from University of Nice Sophia Antipolis at INRIA. He published over 150 scientific peer-reviewed papers, three US patents and is the author of several publications in renowned conferences including ACM SIGCOMM and IEEE INFOCOM. Dali is also a member of the editorial board of the journal on Privacy Enhancing Technologies since 2013.

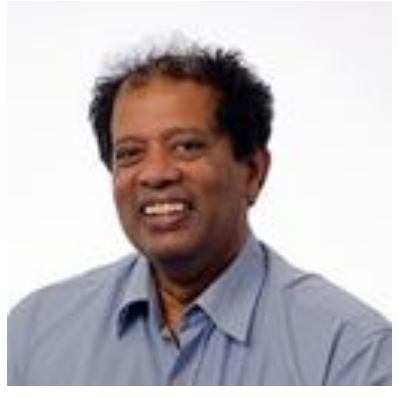

**Aruna Seneviratne** is the foundation Chair in Telecommunications and holds the Mahanakorn Chair of Telecommunications, and the leader of the Networks Research Group at NICTA. His current research interests are in mobile content distributions and preservation of privacy. He received his PhD in electrical engineering from the University of Bath, UK. He has held academic appointments at the University of Bradford, UK, Curtin University, and UTS.